\documentclass[11pt,onecolumn]{IEEEtran}

\usepackage{amsmath,url}
\usepackage{epsfig,amssymb,amsbsy,verbatim,array,enumerate}
\usepackage{pstricks,psfrag,theorem,cite,paralist,subfigure,empheq}

\let\intern=\iftrue

\newcommand{\argmin}{\operatornamewithlimits{arg\ min}}
\def\figref#1{Fig.\,\ref{#1}}%
\def\E{\mathbb{E}}
\def\P{\mathbb{P}}
\def\R{\mathbb{R}}

\def\N{\mathbb{N}}

\def\calR{\mathcal{R}}
\def\ie{{\em i.e.}}
\def\eg{{\em e.g.}}

\def\var{\operatorname{var}}

\def\sir{\mathsf{SIR}}

\def\tx{\mathsf{tx}}

\def\sinc{\operatorname{sinc}}
\def\dd{\mathrm{d}}

\def\one{\mathbf{1}}
\def\ps{p_{\rm s}}
\def\Ps{P_{\rm s}}

\def\misr{\mathsf{MISR}}
\def\B{{\rm B}}

\def\mh#1{\shade{\text{#1}}}

\newtheorem{fact}{Fact}
\newtheorem{theorem}{Theorem}

\newtheorem{corollary}{Corollary}
\newtheorem{definition}{Definition}

\newlength{\figwidth}


\usepackage{xcolor,tikz}

\newcommand{\shade}[1]{%
  \colorbox{blue!30}{$\displaystyle#1$}}

\let\arxiv\iftrue

\begin{document}
\title{The Meta Distribution of the SIR\\in Poisson Bipolar and Cellular Networks}
\arxiv\author{Martin Haenggi\\Dept.~of Electrical Engineering\\University of Notre Dame, IN 46556, USA}
\else\author{Martin Haenggi, \IEEEmembership{Fellow, IEEE}
\thanks{Manuscript date \today. The support of the NSF (grant CCF 1216407) is gratefully acknowledged.}
}\fi

\maketitle
\begin{abstract}
The calculation of the SIR distribution at the typical receiver (or, equivalently, the
success probability of transmissions over the typical link) in Poisson bipolar
and cellular networks with Rayleigh fading is relatively straightforward,
but it only provides limited information on the success probabilities
of the individual links.

This paper introduces the notion of the {\em meta distribution} of the SIR, which is
the distribution of the conditional success probability $\Ps$ given the point process, and
provides bounds, an exact analytical expression, and a simple approximation for it.
The meta distribution provides fine-grained information on the SIR and answers questions
such as ``What fraction of users in a Poisson cellular network
achieve 90\% link reliability if the required SIR is 5 dB?".

Interestingly, in the bipolar model, if the transmit probability $p$ is reduced while increasing
the network density $\lambda$ such that the density of concurrent transmitters
$\lambda p$ stays constant as $p\to 0$, $\Ps$ degenerates to a constant, \ie, all links
have exactly the same success probability in the limit, which is the
one of the typical link. In contrast, in the cellular case, if the interfering base stations are
active independently with probability $p$, the variance of $\Ps$ approaches
a non-zero constant when $p$ is reduced to $0$ while keeping the mean success
probability constant.
\end{abstract}
\section{Introduction}
\subsection{Motivation} 
Stochastic geometry provides the tools to analyze wireless networks with randomly placed nodes.
A key quantity of interest in interference-limited networks is the success probability
$\ps(\theta)\triangleq\P(\sir>\theta)$ of the transmission over the typical link, which
corresponds to the complementary cumulative distribution (ccdf) of the signal-to-interference ratio (SIR).
The calculation of $\ps$ involves {\em spatial averaging}, \ie, the evaluation of a certain expectation over
the point process. While this expected value is certainly important, it does not reveal how concentrated
the link success probabilities are. For example, in one network model, all links (or users) could have success
probabilities between $0.85$ and $0.95$, while in another, some links may have $0.5$ and some
may have $0.99$. In both cases, we may find $\ps=0.9$, but the performances of the two networks
in terms of connectivity, end-to-end delay, or quality-of-experience would differ greatly.
Hence it is important to quantify the variability of the link reliabilities around $\ps$.

To this end, our focus in this paper are random variables 
of the form
\begin{equation}
   \Ps(\theta)\triangleq \P(\sir > \theta \mid \Phi, \tx) ,
   \label{Ps}
\end{equation}
where the conditional probability is taken over the fading and the channel access scheme (if random) of
the interferers given the point process and given that the desired transmitter is active.
The goal is to find (or bound) the ccdf of $\Ps$, defined as
\begin{equation}
   \bar F_{\Ps}(x)\triangleq \P^{!{\rm t}}(\Ps(\theta)>x),\quad x\in [0,1],
   \label{FPs}
 \end{equation}
where $\P^{!{\rm t}}$ denotes the reduced Palm measure of the point process, given that
there is an active transmitter at the prescribed location.
Since $\bar F_{\Ps}$ is the (complementary) distribution of a conditional probability, we call it
the {\em meta distribution} of the SIR.
Using this notation, the standard success probability is the mean
\[ \ps(\theta)=\E^{!{\rm t}} (\Ps(\theta))=\int_0^1 \bar F_{\Ps}(x)\dd x .\]
While a direct calculation of the ccdf \eqref{FPs} seems infeasible, we shall see that
the moments of $\Ps(\theta)$ can be expressed in closed-form, which allows the derivation
of an exact analytical expression and simple bounds.
The $b$-th moment of $\Ps(\theta)$ is denoted by $M_b$,
\ie, we define
\[ M_b(\theta)\triangleq \E^{!{\rm t}} (\Ps(\theta)^b)=\int_0^1bx^{b-1} \bar F_{\Ps}(x)\dd x. \]
Hence we have $\ps(\theta)\equiv M_1(\theta)$.

\subsection{Contributions}
The contributions of the paper are:
\begin{itemize}
\item We introduce the {\em meta distribution} of the SIR.
\item We give closed-form expression of the moments $M_b$ for Poisson bipolar networks with ALOHA
and for Poisson cellular networks, both for Rayleigh fading.
\item We provide an analytical expression for the exact meta distribution for the two types of networks.
\item We propose the beta distribution as a highly accurate approximation.
\item 
We show that, remarkably, in the limit of very dense bipolar networks with small transmit probability,
all links have the same success probability. This is not the case in cellular networks with random
(interfering) base station activity, since the variance $M_2-M_1^2$ is bounded away from zero
when the probability of a base station being active goes to $0$.
\item We give the conditions on the SIR threshold $\theta$ and the transmit probability $p$ for
a finite mean local delay.
\end{itemize}

\subsection{Related work}
The calculation of the (mean) success probability $\ps(\theta)$ in Poisson bipolar networks
is provided in \cite{net:Baccelli06} but can be traced back to \cite{net:Zorzi95}. 
In \cite{net:Ganti10asilomar}, the moments $M_b$ of the link
success probabilities are calculated under the assumption of no MAC scheme
(i.e., all nodes always transmit), and bounds on the distribution are obtained.

For Poisson cellular models, where the typical user is associated with the nearest base station
(strongest base station on average),
the result was derived in \cite{net:Andrews11tcom} and extended to the multi-tier Poisson case
(HIP model) in \cite{net:Nigam14tcom}. 

The joint success probability of multiple transmissions in Poisson bipolar networks is
calculated in \cite{net:Haenggi13twc}. Similarly, 
\cite{net:Zhang14twc} determined the joint success probabilities of multiple transmissions
(or transmissions over multiple resource blocks) for Poisson cellular networks.
As we shall see, these joint probabilities are related to the integer moments $M_k$ of the
conditional success probabilities. 

\subsection{The meta distribution}
In this section, we formally introduce the concept of a {\em meta distribution}, which is the
distribution of the conditional distribution $\Ps$.
\begin{definition}[Meta distribution]
The meta distribution of the SIR is the two-parameter distribution function
\[ \bar F(\theta,x)\triangleq \bar F_{\Ps}(\theta,x)=\P^{!{\rm t}}(\Ps(\theta)>x),\quad \theta\in\R^+,\:x\in [0,1]. \]
\end{definition}
We have $\bar F(0,x)=1$ for $x<1$, $\lim_{\theta\to\infty} \bar F(\theta,x)=0$ for $x>0$, $\bar F(\theta,0)=1$, and
$\bar F(\theta,1)=0$.
For fixed $\theta$, it is a standard ccdf and yields the probability that the typical link or user
achieves an SIR of $\theta$ or, equivalently, the fraction of links or users (assuming a uniform
user distribution) that achieve this SIR. Generally, it yields the fraction of links or users that achieve
an SIR of $\theta$ with probability at least $x$.

In the next two sections, we will calculate the meta distribution and bounds for Poisson bipolar
and cellular networks, respectively.

\section{Poisson Bipolar Networks}
\subsection{System Model}
We consider the {\em Poisson bipolar model} \cite[Def.~5.8]{net:Haenggi12book},
where the (potential) transmitters form a Poisson point process (PPP) $\Phi$ of intensity
$\lambda$ and each one has a dedicated receiver at distance $R$ in a random
orientation. In each time slot, nodes in $\Phi$ independently transmit
with probability $p$, and all channels are subject to Rayleigh fading.

We use the standard path loss model with exponent $\alpha$, define
$\delta\triangleq 2/\alpha$, and we let
$C\triangleq \lambda \pi R^2\Gamma(1-\delta)\Gamma(1+\delta)$ be a coefficient
that does not depend on $\theta$.
The success probability of the typical link is well known, see, \eg, \cite{net:Baccelli06,net:Haenggi08now,net:Haenggi12book},
and can be expressed as
\[ \ps(\theta)\triangleq\P^{!{\rm t}}(\sir>\theta)=M_1(\theta)=e^{-C\theta^\delta p} .\]

Due to the ergodicity of the PPP, the ccdf of $\Ps$ can be alternatively written as
the limit
\[ \bar F_{\Ps}(x)=\lim_{r\to\infty} \frac{1}{\lambda p\pi r^2} \sum_{\substack{y\in\Phi\\\|y\|<r}} 
\one(\P(\sir_{\tilde y}>\theta \mid\Phi)>x) ,\]
where $\tilde y$ is the receiver of transmitter $y$ and $\one(\cdot)$ is the indicator function.
This shows that $\bar F_{\Ps}(x)$ denotes the fraction of links in the network (in each realization of the point process) that,
when scheduled to transmit\footnote{The received signal power is assumed zero if the desired transmitter is
not active, so the SIR is zero in this case.}, have a success probability larger than $x$. 

The link success probabilities for a given realization can also be ``attached" to each point of the transmitter process
$\Phi$ to form a marked point process $\hat\Phi=\{(x_i,\Ps^{x_i})\}$. The meta distribution can then be interpreted
as the mark distribution, parametrized by $\theta$.
Due to the interference correlation \cite{net:Ganti09cl}, the marks of nearby nodes are correlated, hence
$\hat\Phi$ is not an independently marked process.

\begin{figure}
\centerline{\epsfig{file=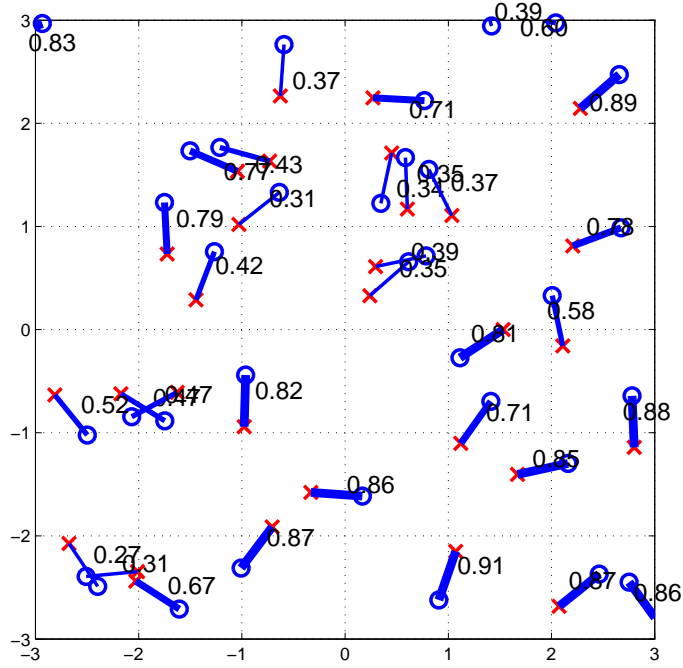,width=\figwidth}}
\caption{Realization of a Poisson bipolar network for $\lambda=1$, $R=1/2$, $p=1/2$, $\theta=1$, $\alpha=4$,
resulting in $\ps=0.54$.
The number next to each link is its success probability (averaged over fading and ALOHA).}
\label{fig:bipolar}
\end{figure}

\figref{fig:bipolar} shows an example realization of a Poisson bipolar network together with the success probabilities
for each link, averaged over the fading and ALOHA.
As expected, links whose receivers are relatively isolated from interfering transmitters have a high success rate,
while those in crowded parts of the network suffer from a low one.

\subsection{Moments}
Let 
\begin{equation}
  D_b(p,\delta)\triangleq \sum_{k=1}^\infty \binom bk\binom{\delta-1}{k-1} p^k ,\quad b\in\mathbb{C} \text{ and }p,\delta\in [0,1].
  \label{dbp}
\end{equation}
For $p=1$,
\[ D_b(1,\delta)=\frac{\Gamma(b+\delta)}{\Gamma(b)\Gamma(1+\delta)} ,\]
which is not defined if $b\in\mathbb{Z}^-$ or $b+\delta\in\mathbb{Z}^-$.
For $\delta\in\{0,1\}$, the function simplifies to $D_b(p,0)=1-(1-p)^b$ and $D_b(p,1)=bp$. 

Alternatively, the function can be
expressed using the Gaussian hypergeometric function $_2F_1$ as
\begin{equation}
 D_b(p,\delta)=pb\:_2F_1(1-b, 1-\delta; 2; p).
\end{equation}
\begin{theorem}[Moments for bipolar network with ALOHA]
\label{thm:moments}
Given that the typical link is active, the moment $M_b$ of the conditional success probability is
\begin{equation}
M_b(\theta)=\exp\left(-C\theta^\delta D_b(p,\delta) \right),\quad b\in\mathbb{C},
\label{moments_b}
\end{equation}
whenever $D_b(p,\delta)$ is defined.
\end{theorem}
{\em Proof: See Appendix A.}

An important and helpful observation in the proof is that the calculation of the $n$-th moment for
$n\in \N$ is the same as that of the joint success probability of $n$ transmissions,
calculated in \cite{net:Haenggi13twc}. In this case, $D_n(p,\delta)$ is given by the finite sum
\[ D_n(p,\delta)=\sum_{k=1}^n \binom nk\binom{\delta-1}{k-1} p^k ,\]
which is a polynomial of degree $n$ in $p$ and degree $n-1$ in $\delta$ and
called the {\em diversity polynomial} in \cite[Def.~1]{net:Haenggi13twc}.

Since \eqref{moments_b} is valid for (essentially) any $b\in\mathbb{C}$, we can use it to obtain 
the $-1$-st moment as
\begin{align}
M_{-1}(\theta)&=\exp(C\theta^\delta p (1-p)^{\delta-1} ) \nonumber \\
&= M_1^{-(1-p)^{\delta-1}},\quad p<1.
\label{moments_m1}
\end{align}

$M_{-1}$ is the mean number of transmission attempts needed to succeed once if
the transmitter is allowed to keep transmitting until success.
This quantity is termed {\em mean local delay} and is calculated in \cite[Lemma 2]{net:Haenggi13tit}.
Noteworthy is the phase transition at $p=1$. For $p=1-\epsilon$, the mean local delay is finite for all
$\epsilon>0$. But if all nodes always transmit, it is infinite.

An interesting question is what happens when $p\to 0$ while the transmitter density
$p\lambda$ (and thus $M_1$) is kept constant. It is answered in the following corollary.

\begin{corollary}[Concentration as \boldmath$p\to 0$]
\label{cor:conc}
Denoting the transmitter density as $\tau\triangleq \lambda p$ and keeping it (and thus $M_1$)
fixed while letting $p\to 0$, we have
\[ \lim_{\substack{p\to 0\\\lambda p=\tau}} \Ps(\theta) = \ps(\theta) \]
in mean square (and probability and distribution). 
\end{corollary}
\begin{IEEEproof}
From \eqref{moments_b}, the second moment is  
\[ M_2(\theta)=e^{-C\theta^\delta(2p+(\delta-1)p^2)} ,\]
and the variance, expressed in terms of $M_1$ (which is kept constant), is
\begin{equation}
   \var \Ps(\theta)=M_1^2(M_1^{p(\delta-1)}-1).  
   \label{var}
\end{equation}
It follows that 
\[ \lim_{\substack{p\to 0\\\lambda p=\tau}} \var\Ps(\theta) = 0. \]
\end{IEEEproof}

So if $C\theta^\delta p$ is kept constant, the variance can be adjusted by changing $p$.
For example, if $C=1/(10p\theta^\delta)$, $M_1=e^{-1/10}\approx 0.9$, and the variance can be reduced to $0$
by letting $p\to 0$. So, counterintuitively, a small $p$ {\em decreases} the variance and, in the limit,
{\em all links in the network have exactly the same success probability}.

More precisely, the variance is proportional to $p$ for small $p$ if $M_1$ is kept constant:
\[ \var \Ps(\theta) \sim -M_1^2\log(M_1) (1-\delta) p,\quad p\to 0 .\]

The next result provides tight bounds on the moments if $p=1$ for $b\in\mathbb{R}^+$.
$'\!\!\lesssim'$ and $'\!\!\gtrsim'$ indicate upper bound and lower bounds with asymptotic equality (here as $b\to\infty$),
respectively.
\begin{corollary}[Bounds on moments for \boldmath$p=1$]
For $b>0$,
\begin{equation}
M_b=M_1^{\frac{\Gamma(b+\delta)}{\Gamma(1+\delta)\Gamma(b)}} \gtrsim \exp(-C \theta^\delta b^\delta) ,
\label{mb_p1_lower}
\end{equation}
for $b\geq 1$,
\begin{equation}
M_b \leq M_1^{b^\delta},
\label{mb_p1_upper}
\end{equation}
and for $0<b<1$,
\begin{equation}
M_b > M_1^{b^\delta}.
\label{mb_p1_lower2}
\end{equation}
\end{corollary}
\begin{IEEEproof}
The lower bound \eqref{mb_p1_lower} follows from \eqref{moments_b} by setting $p=1$ and the
asymptotic bound $\Gamma(b+\delta)/\Gamma(b) \lesssim b^\delta$ for $b>0$.
Conversely, 
$\Gamma(b+\delta)/\Gamma(b) \geq b^\delta\Gamma(1+\delta)$ for all $b\geq 1$,
which yields the upper bound \eqref{mb_p1_upper}:
\begin{align*}
  M_b &\leq \exp(-C b^\delta\Gamma(1+\delta))=M_1^{b^\delta},\quad b\geq 1.
\end{align*}
For $b<1$, $\Gamma(b+\delta)/\Gamma(b) < b^\delta\Gamma(1+\delta)$, and
the direction of the inequality is reversed, yielding \eqref{mb_p1_lower2}.
\end{IEEEproof}
The third bound is tighter than the first one in the regime where it is valid.
Further, since
\[ M_1^{b^\delta}=\exp\left(-C (b\theta)^\delta\right), \]
the $b$-th moment is bounded by the first moment evaluated at $b\theta$, \ie,
\[ M_b(\theta)\leq M_1(b\theta) ,\quad b\geq 1,\]
and vice versa if $b<1$.

\subsection{Exact expression}
An exact integral expression can be obtained from the purely imaginary moments $M_{jt}$, $t\in \R$, $j\triangleq\sqrt{-1}$.

\begin{corollary}[Exact integral expression]
\label{cor:gil}
The meta distribution is given by 
\begin{equation}
  \bar F(\theta, x)=\frac12-\frac1\pi \int_0^\infty \frac{e^{-C\theta^\delta \Re(D_{jt})}\sin(t\log x+C\theta^\delta\Im(D_{jt}))}
       {t}\dd t ,
\label{gil}
\end{equation}
where $D_{jt}=D_{jt}(p,\delta)$ is given in \eqref{dbp} and $\Re(z)$ and $\Im(z)$ denote the real and imaginary parts of the
complex number $z$, respectively.
\end{corollary}
\begin{IEEEproof}
Let $X\triangleq \log\Ps(\theta)$. The characteristic function of $X$ is
\begin{align*}
\varphi_X(t)&\triangleq \E e^{j tX} = \E(\Ps(\theta)^{jt}) =M_{jt}, \quad t\in\R.
\end{align*}
where $M_{jt}$ is given in \eqref{moments_b}.
Then by the Gil-Pelaez theorem \cite{net:Gil-Pelaez51}, the ccdf of $X$ is given by
\begin{equation}
  \bar F_X(x)=\frac12 + \frac1\pi \int_0^\infty \frac{\Im(e^{-jtx} M_{jt})}{t}\dd t .
\end{equation}
Since $\P(\Ps(\theta)>x)=\P(\log\Ps(\theta)>\log x)$,
\begin{align}
   \bar F_{\Ps}(x)&=\frac12+\frac1\pi \int_0^\infty \frac{\Im(e^{-jt\log x} M_{jt})}{t}\dd t ,
 \end{align}
 and the result follows from Thm.~\ref{thm:moments} and some simplification.
 \end{IEEEproof}

Since $|M_{jt}|$ essentially decreases exponentially with $t$, this integral can
be evaluated very efficiently. The curve marked with $\circ$ in \figref{fig:conv_bound3} shows
the exact meta distribution $\bar F(1,x)$ for $\lambda p=1/4$ with different values of $\lambda$ and $p$.
As predicted by Cor.~\ref{cor:conc}, the variance of $\Ps$ is reduced when $p$ is smaller.
Next we will derive the bounds also shown in the figure.

\subsection{Classical bounds on the meta distribution}
Simple bounds on the meta distribution can be established using classical methods.

\begin{figure}
\parbox[c]{.5\textwidth}{%
\centerline{\subfigure[$\lambda=1$, $p=1/4$, and $\var(\Ps)=0.0212$.]
{\epsfig{file=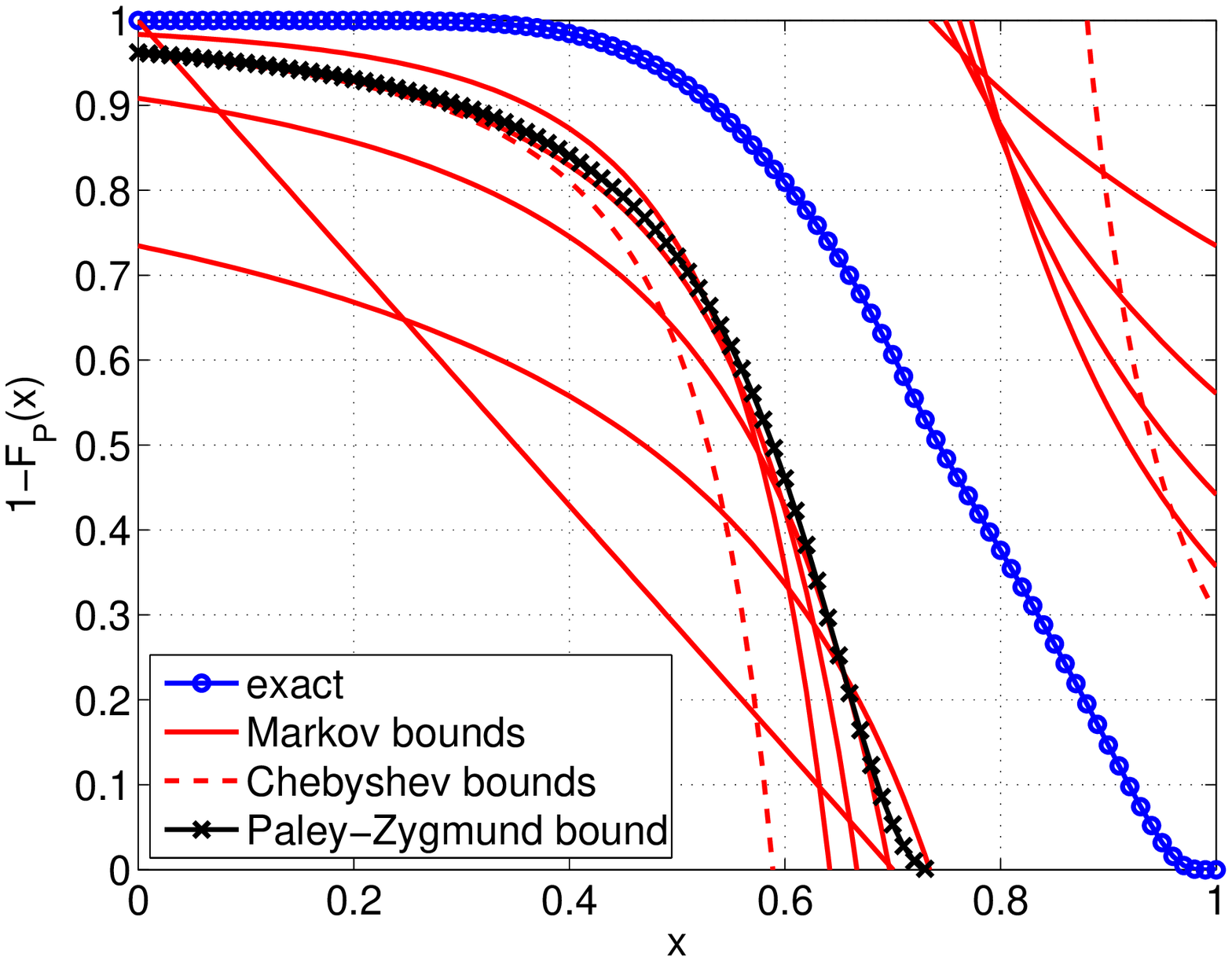,width=.48\textwidth}}}} 
\parbox[c]{.5\textwidth}{%
\centerline{\subfigure[$\lambda=5$, $p=1/20$, and $\var(\Ps)=0.00418$]
{\epsfig{file=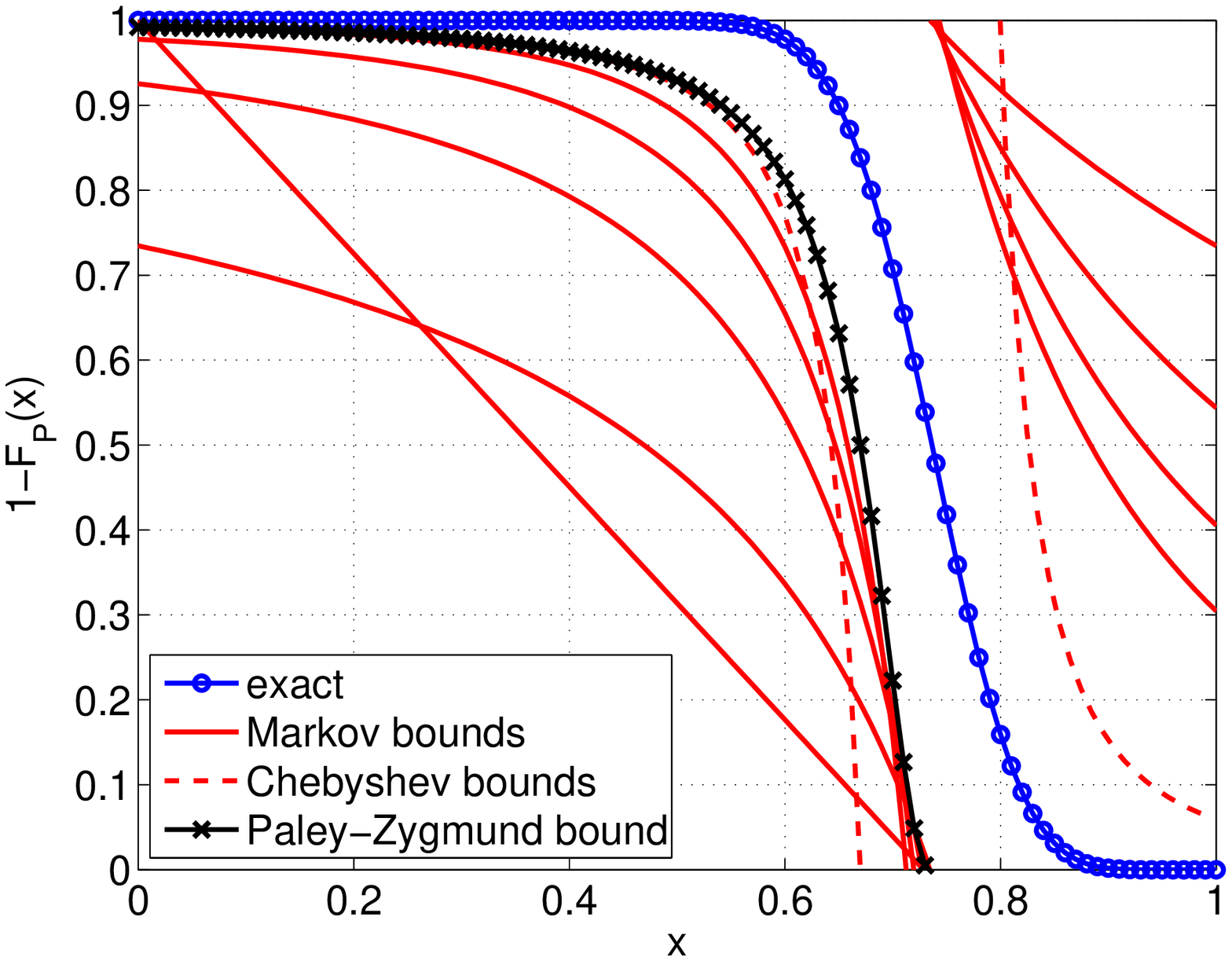,width=.48\textwidth}}}} 
\caption{The exact meta distribution \eqref{gil} and the Markov bounds \eqref{markov} for $b\in[4]$, \eqref{cheby1}, and \eqref{cheby2} for
$\alpha=4$, $\theta=1$, $R=1/2$,
and $\lambda p=1/4$. The resulting mean success probability is $p_s=M_1=0.735$. The variance depends on the
values of $p$ and $\lambda$; it is proportional to $p$ for small $p$.}
\label{fig:conv_bound3}
\end{figure}

\begin{corollary}[Markov and Chebyshev bounds]
\label{cor:classical}
For $x\in [0,1]$, the meta distribution is bounded as
\begin{equation}
  1-\frac{\E^{!t}((1-\Ps(\theta))^b)}{(1-x)^b} < \bar F(\theta,x) \leq \frac{M_b}{x^b} , \quad b>0.
  \label{markov}
\end{equation}
Let $V\triangleq \var \Ps(\theta)=M_2-M_1^2$.
For $x<M_1$,
\begin{equation}
 \bar F_{\Ps}(x)  > 1-\frac{V}{(x-M_1)^2} ,
  \label{cheby1}
\end{equation}
while for $x>M_1$,
\begin{equation}
\bar F_{\Ps}(x) \leq  \frac{V}{(x-M_1)^2}.
\label{cheby2}
\end{equation}
Lastly,
\begin{equation}
 \bar F_{\Ps}(xM_1)\geq \frac{(1-x)^2}{1-M_1^{p(1-\delta)}+(1-x)^2},\quad x\in (0,1).
 \label{p-z}
 \end{equation}
\end{corollary}
\begin{IEEEproof}
\eqref{markov} follows from Markov's inequality, while \eqref{cheby1} and \eqref{cheby2} follow
from Chebyshev's inequality. The lower bound \eqref{p-z} is the Paley-Zygmund (or Cauchy-Schwarz) bound.
\end{IEEEproof}
For the lower (or reverse) Markov bound in \eqref{markov}, the integer moments of $1-\Ps(\theta)$
are easily found using binomial expansion.
For $b=-1$, the Markov inequality also yields the lower bound
$\bar F_{\Ps}(x)\geq 1-xM_{-1}$, where $M_{-1}$ is given in \eqref{moments_m1}.

These bounds are illustrated in the two plots in \figref{fig:conv_bound3}.
For the Markov bounds, the four lower and upper bounds correspond to $b=1,2,3,4$.
It is apparent that the variance decreases with decreasing $p$ and that the bounds get 
tighter also.

Written differently, \eqref{cheby1} and \eqref{cheby2} state that
\[ \bar F_{\Ps}(qM_1) > 1-\frac{M_1^{\delta-1}-1}{(1-q)^2} ,\quad 0<q<1,\] 
and
\[ \bar F_{\Ps}(q M_1) \leq \frac{M_1^{\delta-1}-1}{(1-q)^2} ,\quad 1<q<M_1^{-1}.\]
The upper bound is useful for small $M_1$, while the lower bound is useful for $M_1\approx 1$.

So as $p\to 0$, $\P(\Ps(\theta)\geq xM_1)\to 1$ $\forall x\in (0,1)$, in accordance with Cor.~\ref{cor:conc}.

The Paley-Zygmund bound is useful to bound the fraction of links that has at least a certain fraction of the
average performance. For example, the fraction of links having better than half the
average reliability is lower bounded as
\[ \P^{!{\rm t}}(\Ps(\theta)\geq M_1/2)\geq \frac{1/4}{5/4-M_1^{p(1-\delta)}} .\]
As $p\to 0$, the lower bound approaches $1$, again
as expected from the concentration result in Cor.~\ref{cor:conc}.

\subsection{Best bounds given four moments}
Here we establish the tightest possible lower and upper bounds on the distribution given the first four moments.
Generally, this problem can be formulated as follows.
Letting $\mathcal{M}_k$ be the class of distributions (cdfs) with moments $M_1,\ldots, M_k$, we would like to find
\[ L(x)\triangleq\min_{F\in \mathcal{M}_k} F(x),\quad x\in (0,1) \]
and
\[ U(x)\triangleq\max_{F\in \mathcal{M}_k} F(x),\quad x\in (0,1). \]
So for each $x$ in the support of the distribution, we would like to find the minimum and maximum over
all distributions with the prescribed $k$ moments.
To find $L$ and $U$ for $k=4$, we are applying the method from \cite{net:Racz06mcm}.
It determines
the best lower and upper bounds
\[ L(x) \leq F_Y(x) \leq U(x) \]
given the four moments $\E(Y^k)$, $k\in [4]$,
for a general continuous random variable $Y$.

To bound the cdf $F_Y(x)$ at a target value $x$, first the moments are calculated for the random
variable shifted by $x$ so that the new target location is $0$, \ie,
\begin{align*}
  m_i(x)&\triangleq \int_{0}^{1} (y-x)^i \, \dd F_Y(y) \\
  &= \sum_{k=0}^i \binom ik (-x)^{i-k} \E(Y^k) ,\quad x\in [0,1].
\end{align*}
Using these shifted means, following  \cite{net:Racz06mcm}, 
we define (omitting the dependence on $x$ of the shifted moments to avoid overly cumbrous notation)
\begin{align*}
q(x)&\triangleq \sqrt{(-m_2m_3+m_1m_4)^2-4(m_2^2-m_1m_3)(m_3^2-m_2m_4)} \\
    p_0(x)&\triangleq \frac{-m_2^3+2m_1m_2m_3-m_3^2-m_1^2m_4+m_2m_4} {m_2m_4-m_3^2} \\
    y_1(x)&\triangleq \frac{m_2m_3-m_1m_4-q(x)}{2(m_2^2-m_1m_3)} \\
    y_2(x)&\triangleq \frac{m_2m_3-m_1m_4+q(x)}{2(m_2^2-m_1m_3)} \\
    p_2(x)&\triangleq -\frac{m_2^2-m_1m_3}{q(x)}\left(-m_1-\frac{(m_2^3-2m_1m_2m_3+m_1^2m_4)(-m_2m_3+m_1m_4+q(x)}{2(m_2^2-m_1m_3)(-m_3^2+m_2m_4)}\right) \\
    p_1(x)&\triangleq 1-p_0(x)-p_2(x),
 \end{align*}
 and the bounds follow as
 \begin{equation}
  L(x)=\begin{cases} p_1(x)+p_2(x) & \text{if } y_1(x)<0, \: y_2(x) < 0 \\
       p_1(x) & \text{if } y_1(x)<0,\: y_2(x)>0 \\
       0 & \text{if } y_1(x)>0,\: y_2(x)>0
       \end{cases} 
    \label{best_ub}
 \end{equation}
 \begin{equation}
   U(x)=\begin{cases} 1 & \text{if } y_1(x)<0, \: y_2(x) < 0 \\
       p_0(x)+p_1(x) & \text{if } y_1(x)<0,\: y_2(x)>0 \\
       p_0(x) & \text{if } y_1(x)>0,\: y_2(x)>0
       \end{cases} 
     \label{best_lb}
 \end{equation}
Since $q(x)>0$, it is not possible that $y_1(x)>0$ and $y_2(x)<0$.

In our application $Y=\Ps(\theta)$, $\E(Y^k)=M_k$, and
since we are working with ccdfs, we have
\[ 1-U(x) \leq \bar F(\theta,x) \leq 1-L(x). \]

\begin{figure}
\parbox[c]{.5\textwidth}{%
\centerline{\subfigure[$\lambda=1$ $\Rightarrow$ $\ps=0.54$, $\var(\Ps)=0.049$.]
{\epsfig{file=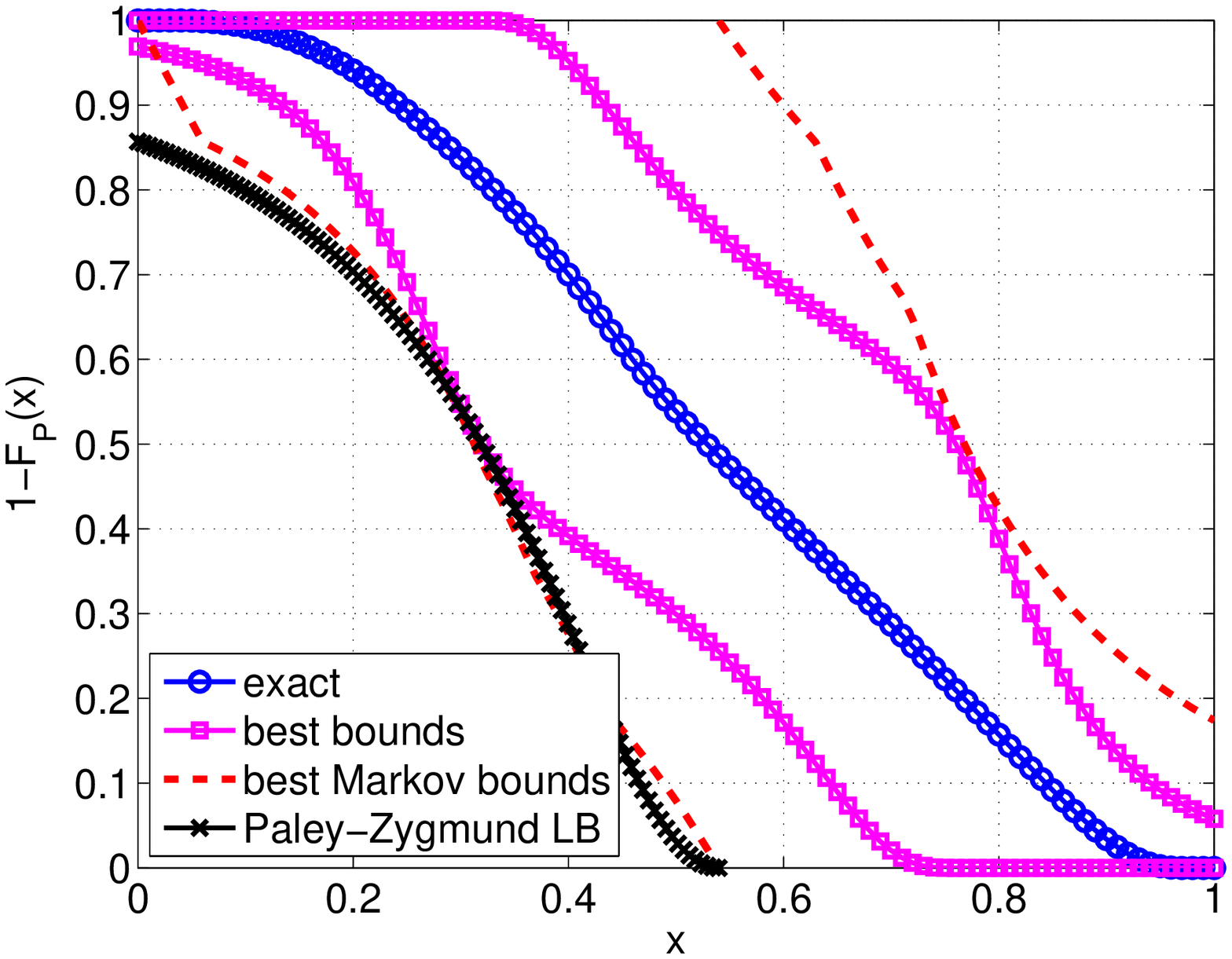,width=.48\textwidth}}}} 
\parbox[c]{.5\textwidth}{%
\centerline{\subfigure[$\lambda=1/5$ $\Rightarrow$ $\ps=0.88$, $\var(\Ps)=0.024$]
{\epsfig{file=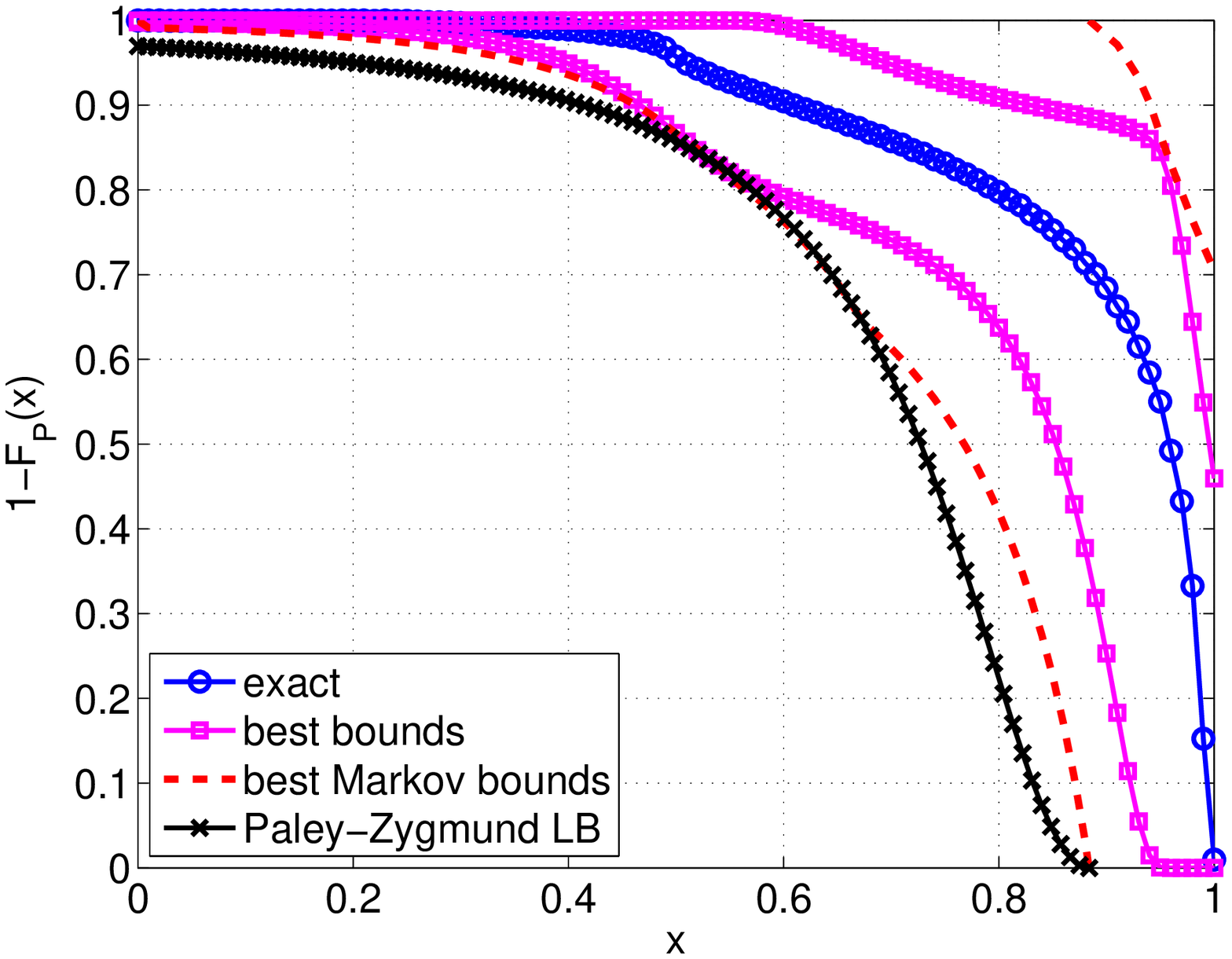,width=.48\textwidth}}}} 
\caption{The exact meta distribution \eqref{gil}, the best Markov bounds \eqref{markov} for $b\in[4]$, and the best overall bounds
per \eqref{best_ub} and \eqref{best_lb} (given the first four moments) for
$\alpha=4$, $\theta=1$, $R=1/2$, and $p=1/2$. The reduction of $\lambda$ from $1$ to $1/5$ results in a reduction
of the variance of only $1/2$, since $p$ stays the same.}
\label{fig:all_bounds}
\end{figure}

\figref{fig:all_bounds} shows these best bounds, 
together with the lower and upper envelopes of the Markov upper and lower bounds for $b\in[4]$ and
the Paley-Zygmund lower bound. In some intervals, the classical bounds are near-optimum, while in others,
 the best bounds are significantly tighter.

The method in  \cite{net:Racz06mcm} is not restricted to four moments, but it is considerably more
tedious to apply if more moments are considered.

\subsection{Approximation with beta distribution}
Since $\Ps(\theta)$ is supported on $[0,1]$, a natural choice for a simple approximating distribution
is the beta distribution.
The probability density function (pdf) of a beta distributed random variable $X$ with mean $\mu$ is
\[ f_X(x)=\frac{x^{\frac{\mu(\beta+1)-1}{1-\mu}}(1-x)^{\beta-1}}{\B(\mu\beta/(1-\mu),\beta)} ,\]
where $\B(\cdot,\cdot)$ is the beta function.
The variance is given by
\[ \sigma^2\triangleq \var X=\frac{\mu(1-\mu)^2}{\beta+1-\mu} .\]
Matching mean and variance $\sigma^2$ yields $\mu=M_1$ and
\[ \beta=\frac{\mu(1-\mu)^2}{\sigma^2}-(1-\mu)=\frac{(\mu-M_2)(1-\mu)}{M_2-\mu^2}.\]

\begin{figure}
\parbox[c]{.5\textwidth}{%
\centerline{\subfigure[Parameters from \figref{fig:conv_bound3} (a) and (b).]
{\epsfig{file=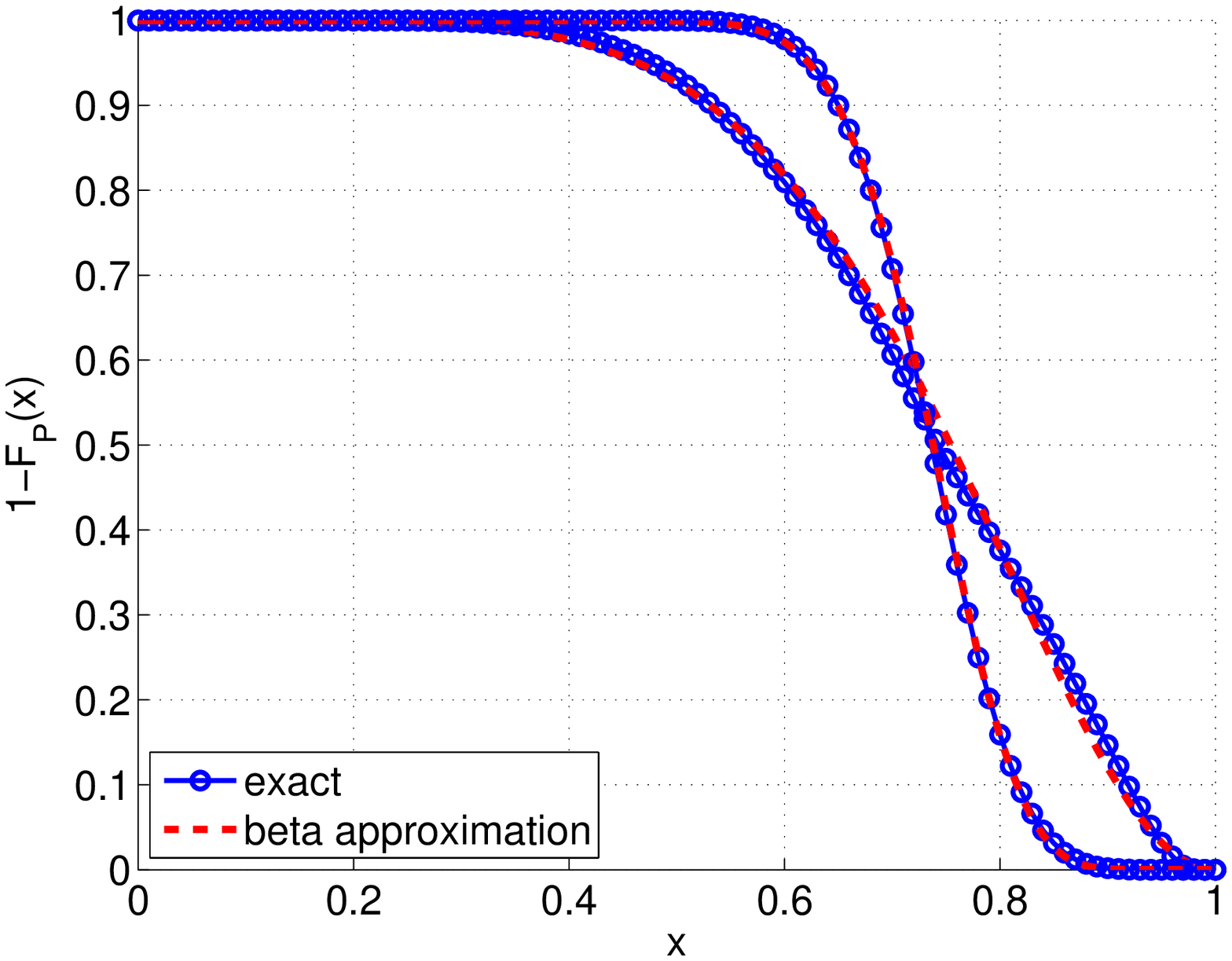,width=.48\textwidth}}}} 
\parbox[c]{.5\textwidth}{%
\centerline{\subfigure[Parameters from \figref{fig:all_bounds} (a) and (b).]
{\epsfig{file=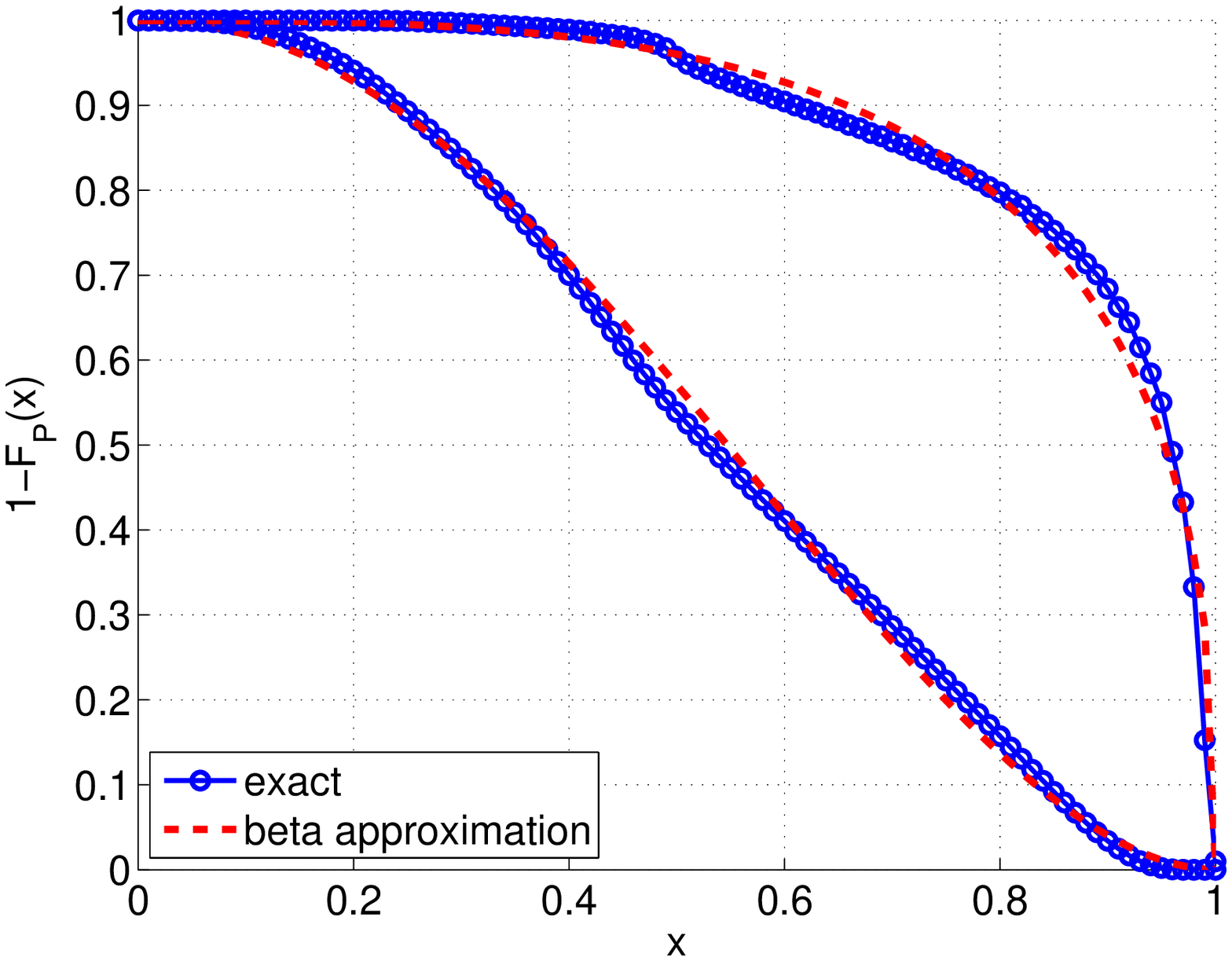,width=.48\textwidth}}}} 
\caption{The exact meta distribution and the beta distribution approximation for the two sets of parameters considered in
the plots of Figs.~\ref{fig:conv_bound3} and \ref{fig:all_bounds}.}
\label{fig:beta_approx}
\end{figure}

As illustrated in \figref{fig:beta_approx} (same parameters as in 
Figs.~\ref{fig:conv_bound3} and \ref{fig:all_bounds}),
the beta distribution provides an excellent match for the distribution of the link success
probabilities, which is also corroborated by the fact that the higher moments $\E(X^k)$ of the
matched beta distribution are very close to $M_k$.
For example, for the parameters
in \figref{fig:conv_bound3}(a), the analytical $-1$-st and $3$-rd through $8$-th moments differ by less
than $3\%$, as shown in Table \ref{table:moments}.
So the skewness and kurtosis and the mean local delay are approximated very accurately also.

\begin{table}
\begin{center}
\begin{tabular}{|l||c|c|c|c|c|c|c|c|c|}
\hline 
 & $k=-1$ & $k=3$ & $k=4$ & $k=5$ & $k=6$ & $k=7$ & $k=8$ \\ \hline\hline
$M_k$ & 1.4278 & 0.4418 &  0.3571 &   0.2947 &   0.2476  &  0.2110 &   0.1820   \\ \hline 
$\E(X^k)$ & 1.4333 & 0.4412 &   0.3555 &   0.2921 &   0.2440 &   0.2066 &   0.1770 \\ \hline 
ratio & 0.9962& 1.0014  &  1.0044  &  1.0090  &  1.0147  &  1.0211 &   1.0280 \\\hline
\end{tabular}
\end{center}
\caption{Comparison of moments $M_k$ and $\E(X^k)$ of the beta approximation for the parameter set in 
\figref{fig:conv_bound3}(a).}
\label{table:moments}
\end{table}

\begin{figure}
\centerline{\epsfig{file=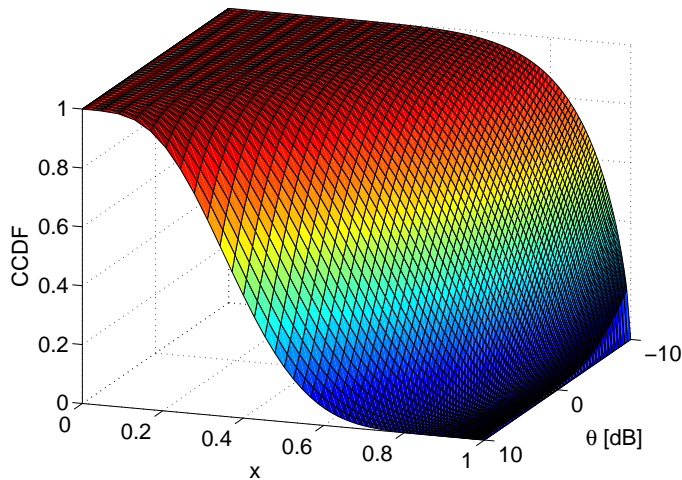,width=\figwidth}}
\caption{Three-dimensional plot of the meta distribution $\bar F(\theta,x)$ for $\lambda=1$, $p=1/4$, $\alpha=4$, and $R=1/2$.}
\label{fig:meta_dist}
\end{figure}

\begin{figure}
\parbox[t]{.5\textwidth}{%
\centerline{\subfigure[Meta distribution for $\theta=-10,-5,0,5,10,15$ dB. The curve for $\theta=0$ dB is marked
with $\circ$.]
{\epsfig{file=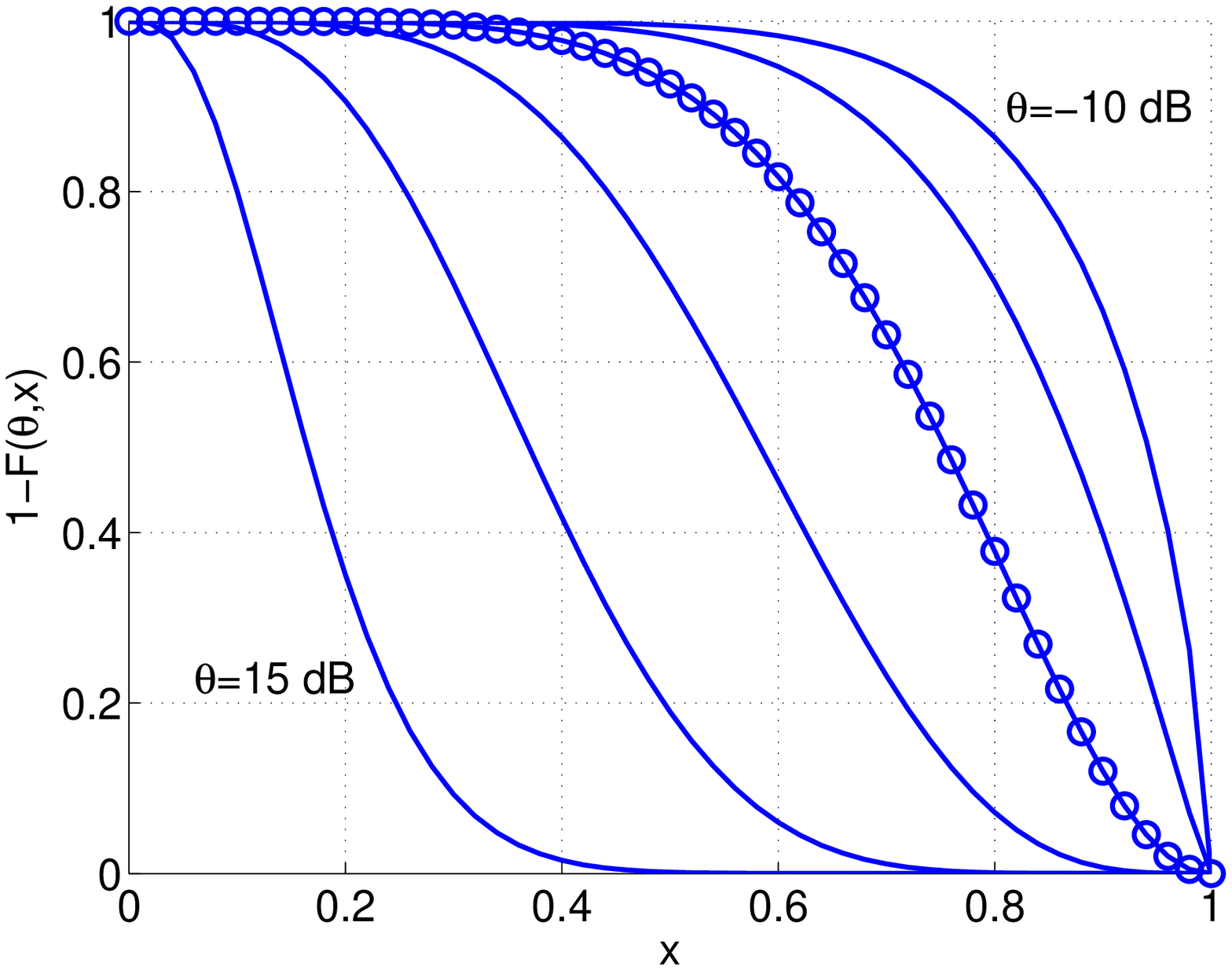,width=.48\textwidth}}}} 
\parbox[t]{.5\textwidth}{%
\centerline{\subfigure[Meta distribution as a function of $\theta$ for $x=0.4,0.5,0.6,0.7,0.8,0.9$.]
{\epsfig{file=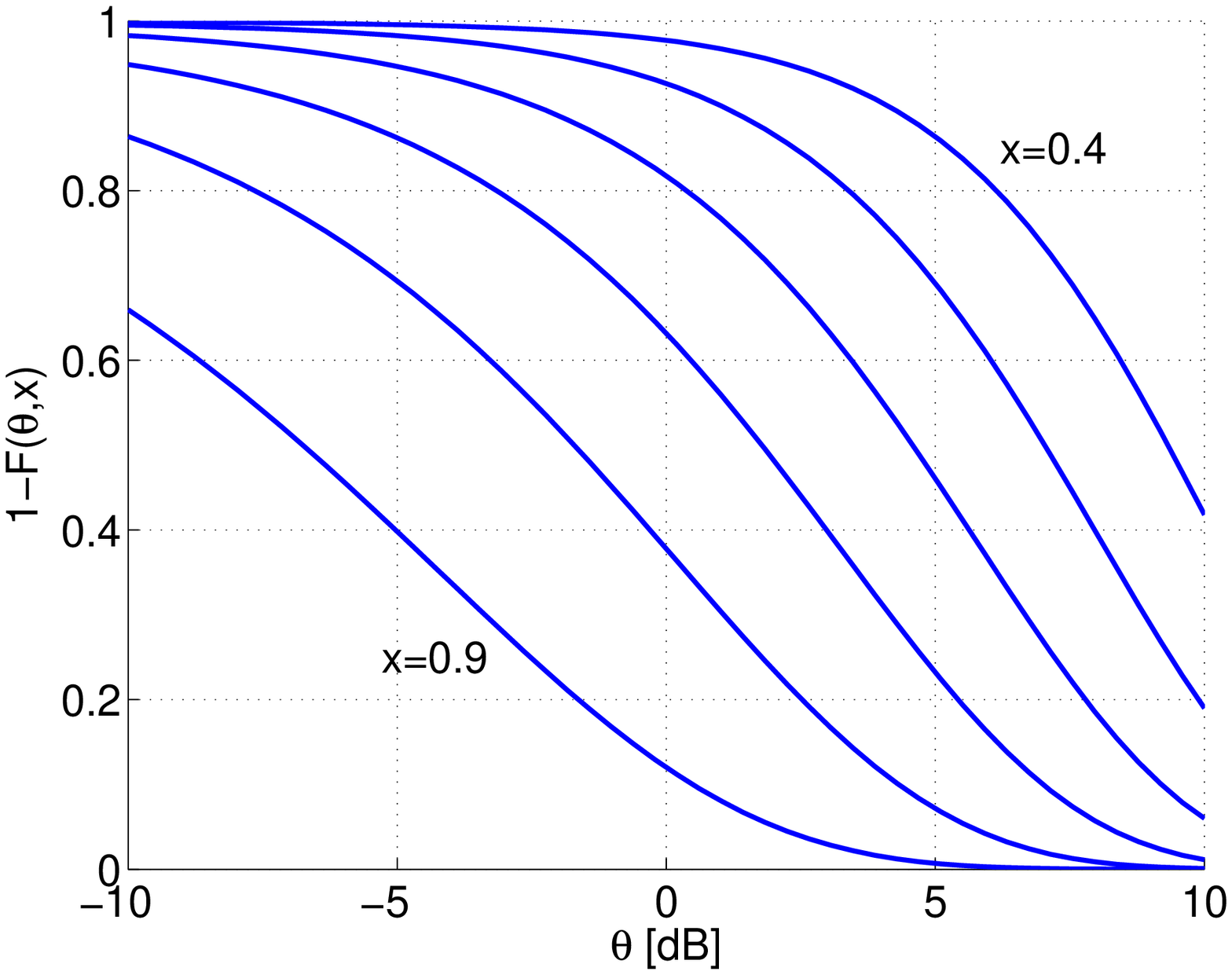,width=.48\textwidth}}}} 
\caption{Cross-sections through the meta distribution along the $x$ and $\theta$ axes for $\lambda=1$, $p=1/4$, $\alpha=4$, $R=1/2$.}
\label{fig:meta_cross}
\end{figure}

\begin{figure}
\centerline{\epsfig{file=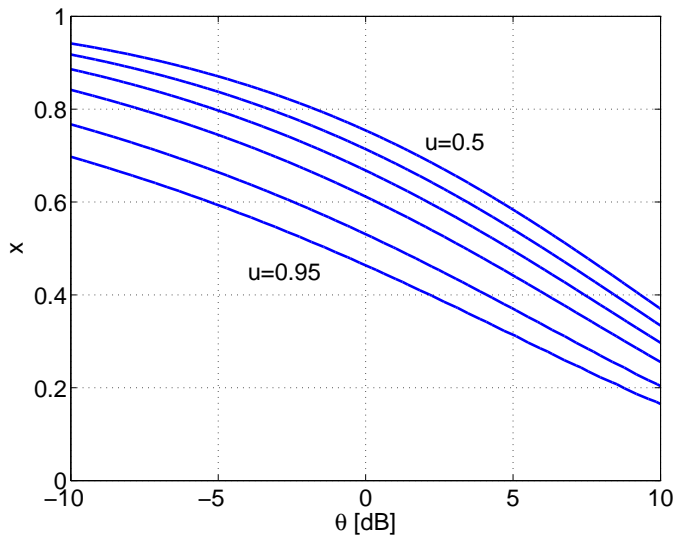,width=\figwidth}}
\caption{Contour plot of meta distribution $\bar F(\theta,x)$ for $\lambda=1$, $p=1/4$, $\alpha=4$, and $R=1/2$.
The values at the curves are $\bar F(\theta,x)=u=0.5,0.6,0.7,0.8,0.9,0.95$ (from top to bottom).}
\label{fig:meta_contour}
\end{figure}

\subsection{Illustrations of the meta distribution}
An illustration of the meta distribution is shown in \figref{fig:meta_dist}. It shows qualitatively that, for the chosen
parameters, most links achieve an SIR of $-10$ dB with probability $80\%$, while an SIR of $10$ is achieved 
with probability $80\%$ by virtually no links.
For quantitative purposes, the cross-sections and contours are more informative, as shown in the next figures.

\figref{fig:meta_cross}(a) enables a more precise statement about the fraction of links achieving an SIR of $-10$ dB with
$80\%$ reliability---it is $0.93$. It also shows that at $\theta=0$ dB, $60\%$ of the
links have a success probability of at least $80\%$.

As a function of $\theta$ for fixed $x$, the value of $\theta$ can be determined such
that at least a fraction $x$ of users have a success probability $p_{\rm min}$. 
For example, \figref{fig:meta_cross}(b) shows that to achieve at least $80\%$ success probability
for $80\%$ of the links, a $\theta$ of at most $-7.6$ dB can be chosen.

The contour plot \figref{fig:meta_contour} visualizes the trade-off between $x$ and $\theta$. 
It shows the combinations $(\theta,x)$ that can be achieved by a certain fraction of links $u$.
For example, the curve for link fraction $u=0.95$ shows that $95\%$ of the links
achieve an SIR of $-5$ dB with probability $0.6$ and an SIR of $5$ dB with probability $0.31$.

Hence the contour plot illustrates and quantifies the trade-off between data rate (as determined by
$\theta$) and reliability (given by the parameter $x$) in bipolar networks.

\section{Poisson Cellular Networks}
\subsection{System model}
\begin{figure}
\centerline{\epsfig{file=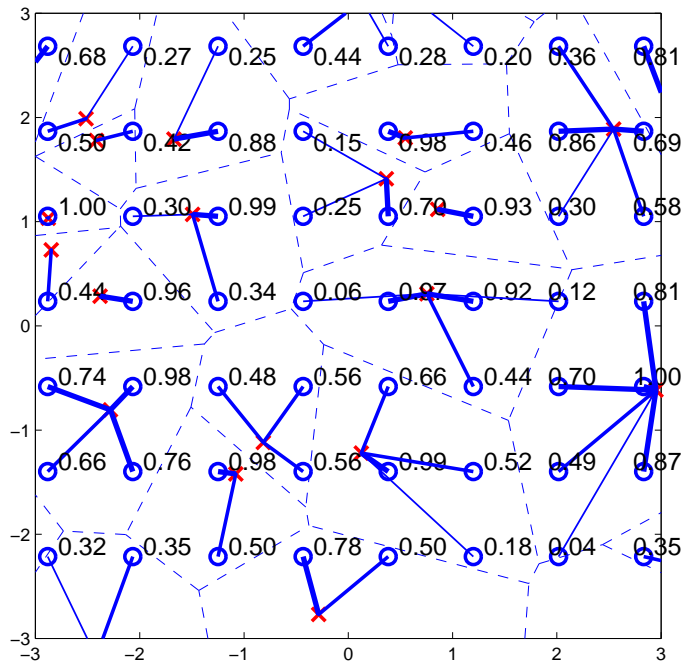,width=\figwidth}}
\caption{Realization of a Poisson cellular network with BS density $\lambda=1$, users forming a square lattice of density 
$\lambda_{\rm u}=3$, $\theta=1$, and $\alpha=4$, resulting in $\ps=0.56$. The BSs are indicated by $\times$ and the users by $\circ$.
The number next to each user is its success probability (averaged over fading) or its mark, and the dashed
lines are the edges of the Voronoi cells of the BS PPP.}
\label{fig:cellular}
\end{figure}

In Poisson cellular networks, base stations (BSs) form a PPP of intensity $\lambda$, while users form a stationary
point process of intensity $\lambda_{\rm u}$. We focus on the downlink and on nearest-BS association,
\ie, each BS serves all the users in its Voronoi cell, and first assume that all BSs are always active.
An example realization where users form a square lattice is shown in \figref{fig:cellular}.

As in the bipolar case, we assume the standard path loss law with path loss exponent $\alpha=2/\delta$ and
Rayleigh fading. The standard (mean) success probability (or SIR distribution) is the success probability of the typical user,
assumed at the origin $o$, which is known from \cite{net:Andrews11tcom} as
\[ \ps(\theta)=\P^o(\sir>\theta)=\frac1{_2F_1(1,-\delta; 1-\delta; -\theta)} .\]
The probability also has a spatial interpretation: for each realization of the BS and user point processes,
it gives the fraction of users achieving an SIR of at least $\theta$ in a given time slot. It depends neither on the
user density nor on the BS density.

Again we define the conditional success probability
\[ \Ps(\theta)\triangleq \P^o(\sir>\theta\mid \Phi) ,\]
which is the probability that the SIR at the origin exceeds $\theta$ given the BS process and given that a user is located at $o$.
The quantity of interest is the meta distribution of the SIR, which is the distribution (ccdf) of $\Ps$:
\[ \bar F(\theta,x)\triangleq \bar F_{\Ps}(x)= \P(\Ps(\theta)>x), \quad \theta\in\mathbb{R}^+,\: x\in [0,1] \]
It gives detailed information about the user experience by providing the fraction of users achieving an SIR
of $\theta$ with reliability at least $x$.

As before, a direct calculation of this meta distribution seems infeasible and we thus focus on the moments
$M_b\triangleq \E(\Ps(\theta)^b)$ first.

\subsection{Moments}

\begin{theorem}[Moments for cellular network]
The moments of the conditional success probability for Poisson cellular networks are given by
\begin{equation}
  M_b=\frac1{_2F_1(b,-\delta; 1-\delta; -\theta)} ,\quad b\in \mathbb{C}.
  \label{mb_cell}
\end{equation}
\end{theorem}
\begin{IEEEproof}
Let $x_0=\argmin\{x\in\Phi\colon \|x\|\}$ be the serving BS of the typical user.
Given the BS process $\Phi$, the success probability is
\begin{align*}
   \Ps(\theta)&= \P\Big(h > \|x_0\|^\alpha \theta \sum_{x\in\Phi\setminus\{x_0\}} h_x\|x\|^{-\alpha} \: \Big| \: \Phi\Big) \\
      &=\prod_{x\in\Phi\setminus\{x_0\}} \frac{1}{1+\theta (\|x_0\|/\|x\|)^\alpha} .
\end{align*}
The $b$-th moment follows as
\begin{equation}
  M_b=\E\prod_{x\in\Phi\setminus\{x_0\}} \frac{1}{(1+\theta (\|x_0\|/\|x\|)^\alpha)^b} .
  \label{mb1}
 \end{equation}
 Instead of calculating this expectation in two steps as usual (first condition on $\|x_0\|$ then take the
expectation w.r.t.~it), we use the recent result \cite[Lemma 1]{net:Ganti15arxiv}, which requires the calculation
of only one finite integral. The lemma gives the pgfl of the {\em relative distance process (RDP)}, defined as
\[ \calR\triangleq \{x\in\Phi\setminus\{x_0\}\colon \|x_0\|/\|x\| \}, \]
when $\Phi$ is a PPP. Since \eqref{mb1}, depends on the BS locations only through the relative distances,
we can directly apply the pgfl of the RDP and obtain
\begin{equation}
 M_b=\frac{1}{1+2\int\limits_0^1 \left(1-\frac1{(1+\theta r^\alpha)^{b}}\right) r^{-3}\dd r} ,
 \label{ps_rdp}
 \end{equation}
 which can be expressed as \eqref{mb_cell}.
\end{IEEEproof}

Sometimes the calculation of the hypergeometric function with negative last argument can cause
numerical problems. In such cases, the alternative form
\[ M_b=\frac{(1+\theta)^b}{_2F_1(b,1; 1-\delta; \theta/(1+\theta))}, \]
obtained through Euler's transformation, is helpful.

For $b=-1$, \eqref{mb_cell} (or \eqref{ps_rdp}---no ``detour" using hypergeometric functions needed in this case) simplifies to
\begin{equation}
   M_{-1}=\frac{1-\delta}{1-\delta(1+\theta)} ,\quad \theta<1/\delta-1.
\label{cell_loc_delay}
\end{equation}
As in the bipolar case, this is the mean local delay if $\theta<1/\delta-1$.
Converseley, if $\theta\geq\alpha/2-1$, the
mean local delay is infinite due to the correlated interference in the system. This {\em phase transition}
in the mean local delay is similar to the one observed in \cite{net:Baccelli10infocom,net:Haenggi13tit,net:Haenggi13twc}
for ad hoc networks. Incidentally, the condition can also be expressed as $\theta\,\misr<1$, where $\misr$ is the 
mean interference-to-signal ratio of the PPP introduced in \cite{net:Haenggi14wcl}.

For $b\in\mathbb{N}$, the moment $M_b$ equals the joint success probability of $b$ transmissions,
which was calculated in \cite[Thm.~2]{net:Zhang14twc} using a different (less direct) method.

\begin{figure}
\centerline{\epsfig{file=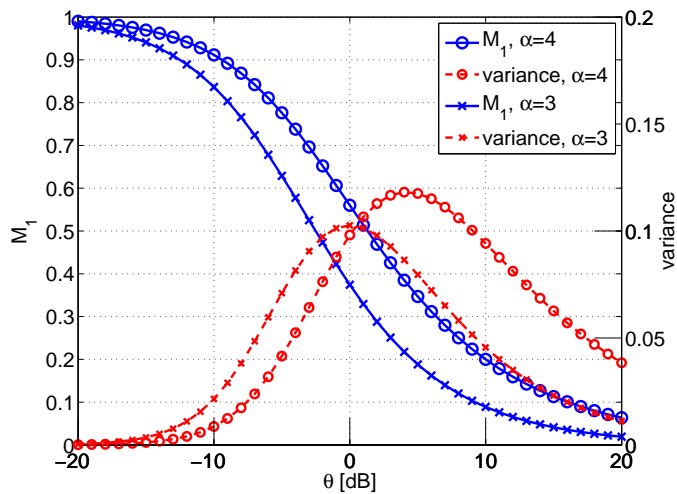,width=\figwidth}}
\caption{Success probability $M_1$ and variance $M_2-M_1^2$ for $\alpha=3$ and $\alpha=4$.}
\label{fig:ps_var}
\end{figure}

\figref{fig:ps_var} shows the standard success probability $M_1=\ps$ and the variance as a function of $\theta$
for $\alpha=3,4$.
Since the variance necessarily tends to zero for both $\theta\to 0$ and $\theta\to\infty$, it assumes a maximum at some
finite value of $\theta$. A numerical evaluation shows that for $\alpha=3$, the variance is maximized
 quite exactly at $\theta=1$, and for both values of $\alpha$, the success probability at which the variance is maximized
 is $\ps=0.38$.

\subsection{Exact expression, bounds, and beta approximation}
As in the bipolar case, we obtain an exact expression for the meta distribution from the Gil-Pelaez theorem.
\begin{corollary}
The SIR meta distribution for Poisson cellular networks is given by
\begin{equation}
 \bar F(\theta,x)=\frac12+\frac1\pi\int_0^\infty \frac{\Im(e^{-jt\log x}M_{jt})}{t}\dd t 
 \label{cell_gil}
\end{equation}
\end{corollary}
Numerical investigations indicate that $|M_{jt}|=\Theta(t^{-1})$, $t\to\infty$, so the integrand decays with $t^{-2}$ and 
the integral can be evaluated efficiently.

\begin{figure}
\parbox[c]{.5\textwidth}{%
\centerline{\subfigure[$\theta=1$ $\Rightarrow$ $\ps=0.56$, $\var(\Ps)=0.098$]
{\epsfig{file=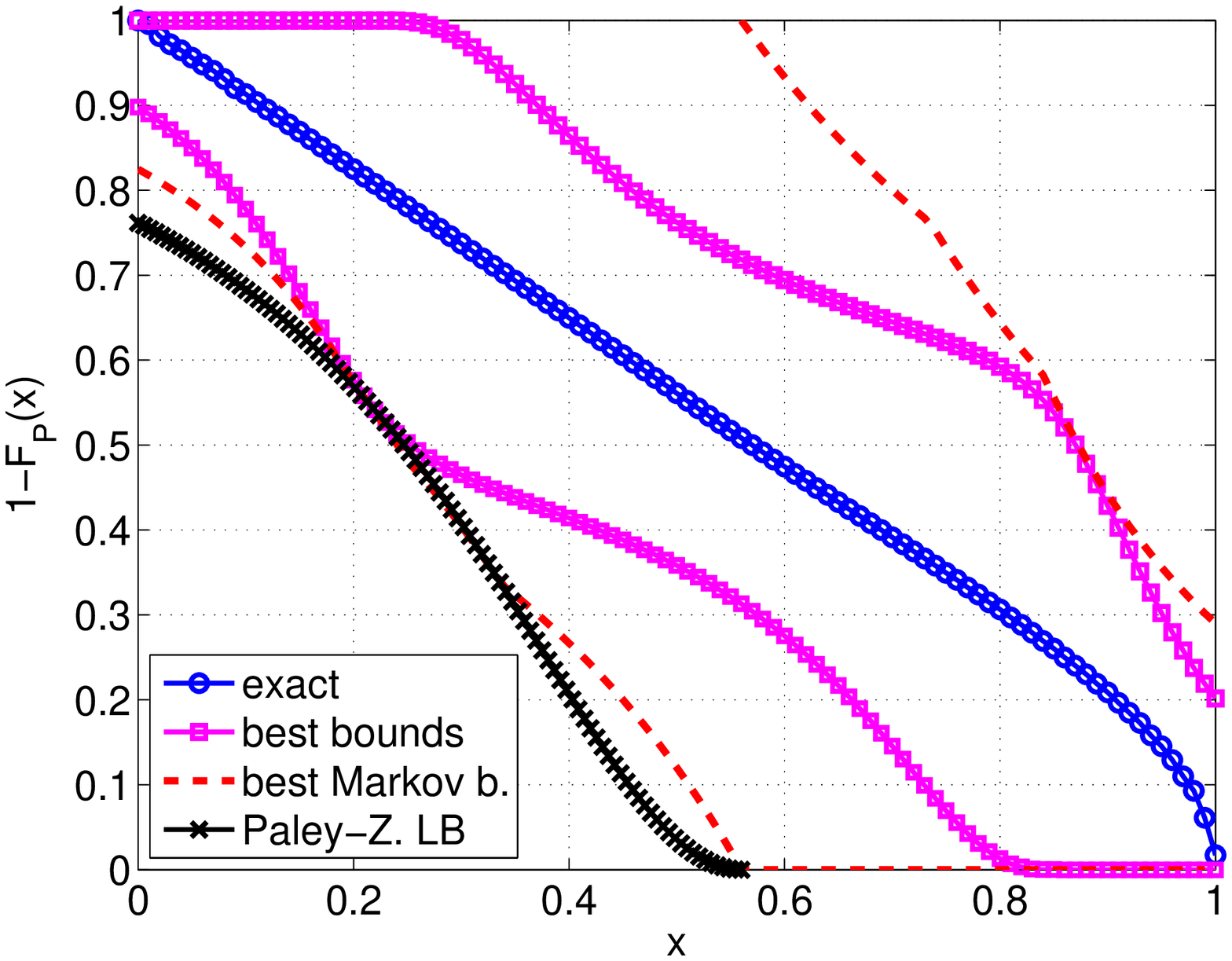,width=.48\textwidth}}}} 
\parbox[c]{.5\textwidth}{%
\centerline{\subfigure[$\theta=1/10$ $\Rightarrow$ $\ps=0.91$, $\var(\Ps)=0.0086$]
{\epsfig{file=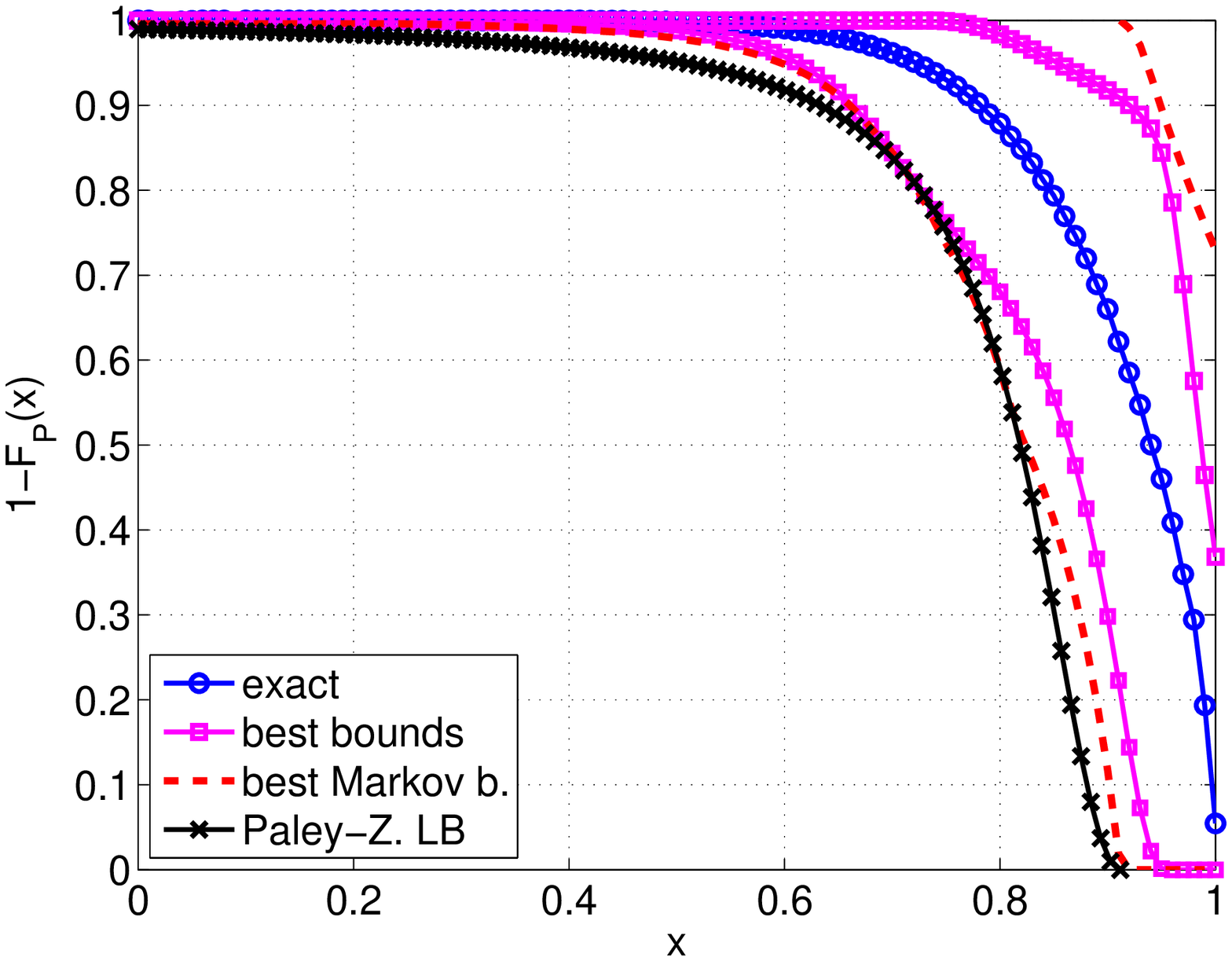,width=.48\textwidth}}}} 
\caption{The exact meta distribution \eqref{cell_gil}, the best Markov bounds \eqref{markov} for $b\in[4]$, the
Paley-Zygmund lower bound, and the best overall bounds (given the first four moments) for
$\alpha=4$.}
\label{fig:all_bounds_cell}
\end{figure}

\figref{fig:all_bounds_cell} shows the exact distribution and the classical and best bounds for $\theta=1$ and $\theta=1/10$, respectively.
Interestingly, the meta distribution $\bar F(1,x)$ has almost constant slope, which means that the user success probabilities
are essentially {\em uniformly distributed} between $0$ and $1$. 

\begin{figure}
\centerline{\epsfig{file=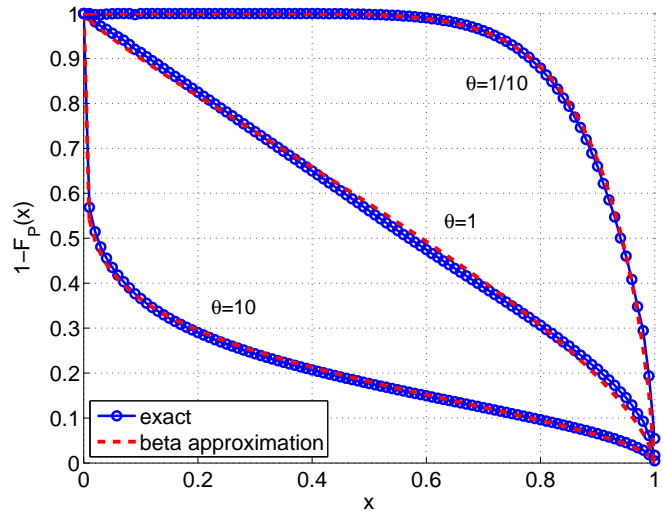,width=\figwidth}}
\caption{Exact ccdf and beta approximation for $\theta=1/10, 1, 10$ for $\alpha=4$.}
\label{fig:beta_cell}
\end{figure}

\figref{fig:beta_cell} shows that the beta approximation provides an excellent fit over a wide
range of $\theta$ values. It also serves as an illustration of the meta distribution showing what 
combinations of reliability $x$ and fraction of users can be achieved for $\theta\in\{-10,0,10\}$ dB.

\begin{figure}
\centerline{\epsfig{file=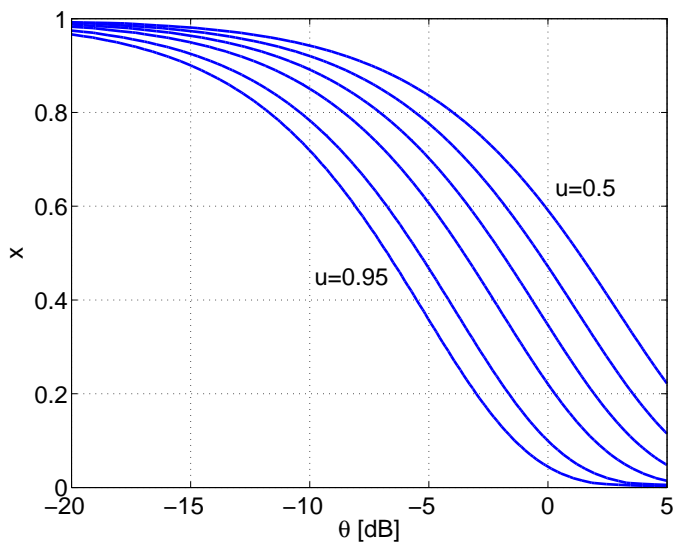,width=\figwidth}}
\caption{Contour plot of meta distribution $\bar F(\theta,x)$ for $\alpha=4$.
The values at the curves are $\bar F(\theta,x)=u=0.5,0.6,0.7,0.8,0.9,0.95$ (from top to bottom).}
\label{fig:meta_contour_cell}
\end{figure}

Lastly, \figref{fig:meta_contour_cell} shows a contour plot of the meta distribution for $\alpha=4$.
An operator who is interested in the performance of the ``5\% user" (the user in the bottom 5-th percentile
in terms of performance) can use the bottom curve, corresponding to $\bar F(\theta,x)=0.95$, to find the performance trade-off
that such a user can achieve. For example, it can achieve an SIR of $-10$ dB with reliability $0.72$
or an SIR of $-4.3$ dB with reliability $0.3$.

\subsection{Effect of random base station activity}
Here we investigate the effect on the meta distribution if interfering BSs were active only
with probability $p$. This is similar to the model studied in \cite[Sec.~VI]{net:Andrews11tcom},
where a frequency reuse parameter $\kappa$ was introduced and each BS is assumed to choose one of
$\kappa$ bands independently at random. Hence the two models are the same if we set $p=\kappa^{-1}$
(apart from the fact that $\kappa\in\mathbb{N}$, whereas no such restriction is imposed on $p^{-1}$).

\begin{theorem}
\label{thm:cell_p}
The $b$-th moment of the success probability in a Poisson cellular network where interfering BSs
are active independently with probability $p$ can be expressed as
 \begin{equation}
 M_b(p)=\left(1-\sum_{k=1}^\infty \binom bk  (-p\theta)^k \frac{\delta}{k-\delta}\, _2F_1(k, k-\delta; k+1-\delta;-\theta)\right)^{-1}.
 \label{mb_cell_p}
 \end{equation}
 \end{theorem}
\begin{IEEEproof}
If interfering BSs are active independently with probability $p$ in each time slot, we have
\begin{align*}
\Ps(\theta) &=\prod_{r\in\calR} \left(\frac{p}{1+\theta r^\alpha}+1-p\right)
\end{align*}
and thus
\begin{align*}
M_b(p)=\E \prod_{r\in\calR} \left(1-\frac{p\theta r^\alpha}{1+\theta r^\alpha}\right)^b.
\end{align*}
Hence we need to modify \eqref{ps_rdp} to
\begin{equation}
 M_b(p)=\frac{1}{1+2\int\limits_0^1 \Big(1-\left(1-\frac{p\theta r^\alpha}{1+\theta r^\alpha}\right)^{b}\Big) r^{-3}\dd r} .
 \label{ps_rdp_p}
 \end{equation}

For general $b\in\mathbb{C}$, letting $x=r^\alpha$, the integral in \eqref{ps_rdp_p} can be
expanded as\footnote{See the appendix, where a similar technique is used.}
\begin{multline}
 \sum_{k=1}^\infty \binom bk \frac{-(-p\theta)^{k}}{\alpha} \int_0^1 \left(\frac{x}{1+\theta x}\right)^k x^{-\delta-1} \dd x =\\
 \sum_{k=1}^\infty \binom bk  \frac{-(-p\theta)^{k}}{k\alpha-2}\, _2F_1(k, k-\delta; k+1-\delta;-\theta) ,
 \end{multline}
 and we obtain the result.
 \end{IEEEproof}
 
  For $b=1$, this yields the success probability
  \begin{align}
   \ps(\theta,p)&=\frac1{1+p\theta\frac{\delta}{1-\delta} \,_2F_1(1,1-\delta;2-\delta,-\theta)}\label{m1_p1}\\
   &=\frac{1}{1-p+p\,_2F_1(1,-\delta; 1-\delta; -\theta)} 
   \label{m1_p}
  \end{align}
  The first expression corresponds to \cite[Eqn.~(19)]{net:Andrews11tcom}, while
  the second one follows from the identity
  \begin{equation}
  \frac{\theta\delta}{1-\delta}\ _2F_1(1,1-\delta; 2-\delta; -\theta)+1\equiv \,_2F_1(1,-\delta; 1-\delta; -\theta). 
  \label{hyper_identity}
  \end{equation}

\begin{figure}
\centerline{\epsfig{file=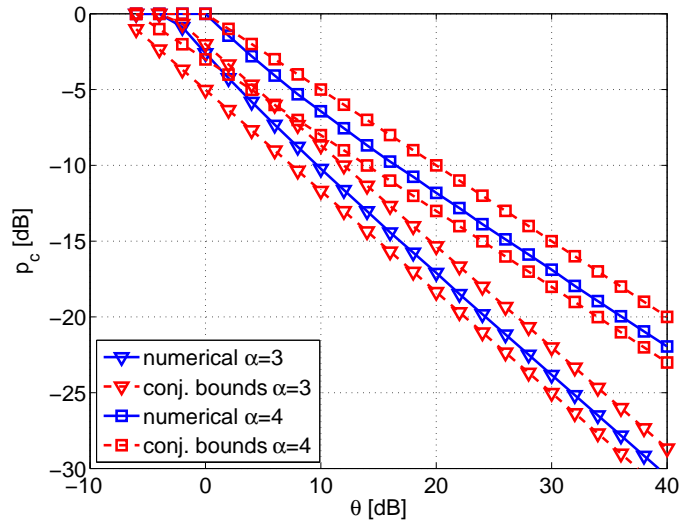,width=\figwidth}}
\caption{Critical probability $p_{\rm c}$ (in dB) for finite mean local delay as a function of $\theta$ for $\alpha=3,4$ and
conjectured lower and upper bounds.}
\label{fig:crit_p_cell}
\end{figure}

For $b=-1$, \eqref{ps_rdp_p} yields
\begin{equation}
  M_{-1}=\frac{1}{1-p \theta\frac{\delta}{1-\delta}\,_2F_1(1,1-\delta;2-\delta,-\theta(1-p))},\quad p\leq p_{\rm c}(\theta). 
  \label{Mm1_p}
\end{equation}
Here $p_{\rm c}(\theta)$ is the critical transmit probability denoting the phase transition from finite to infinite mean
local delay. If $\theta<1/\delta-1$, we know from \eqref{cell_loc_delay} that $p_{\rm c}(\theta)=1$. 
If $p<1$, a larger $\theta$ can be accommodated while maintaining a finite mean local delay.
\figref{fig:crit_p_cell}
shows the critical probability $p_{\rm c}(\theta)$ and two conjectured bounds, which are
$p_{\rm c}(\theta) \geq (\frac{\delta}{1-\delta}\theta)^{-\delta}/2$ and 
$p_{\rm c}(\theta) \leq (\frac{\delta}{1-\delta}\theta)^{-\delta}$.

Next we provide an asymptotic result on the success probability $\ps(p,\theta)$ as $p\to 0$ while keeping
$p\theta^\delta$ constant.

\begin{figure}
\centerline{\epsfig{file=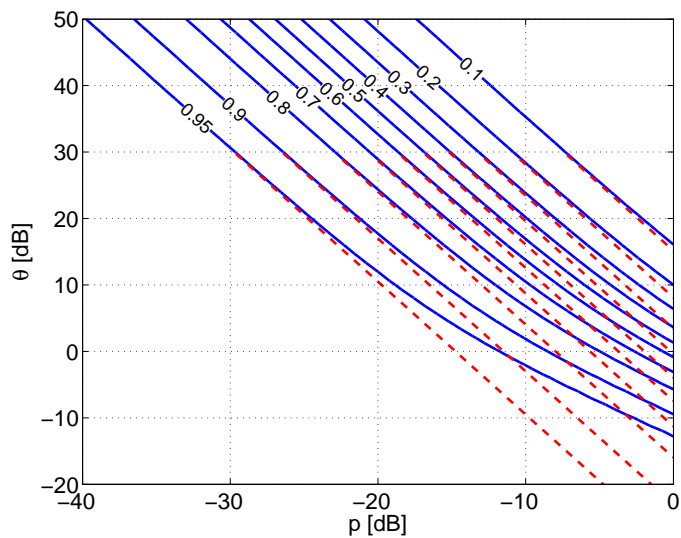,width=\figwidth}}
\caption{Contour plot showing the combinations of $\theta$ and $p$ (in dB) that achieve a given target success probability 
$p_{\rm t}\in\{0.1,0.2,\ldots,0.9,0.95\}$ for $\alpha=4$. The dashed lines are the asymptotes obtained from \eqref{limit_ps}.}
\label{fig:cell_p_contour}
\end{figure}

\begin{corollary}
Let $t=p\theta^\delta$. As $p\to 0$ and $\theta\to\infty$ such that $t$ stays constant,
\begin{equation}
  \ps(\theta,p)\sim \frac{1}{1+p\theta^\delta/\sinc\delta}=\frac{\sinc\delta}{t+\sinc\delta} . 
\label{limit_ps}
\end{equation}
\end{corollary}
\begin{IEEEproof}
From Thm.~4 and Lemma 6 in \cite{net:Ganti15arxiv}, $_2F_1(1,-\delta; 1-\delta; -\theta) \sim \theta^\delta/\sinc\delta$, $\theta\to\infty$.
Inserting this in \eqref{m1_p} and letting $p\to 0$ and $\theta\to\infty$ while keeping $p\theta^\delta$ constant yields the result.
\end{IEEEproof}
The corollary implies that
\[ \ps(\theta,p) \sim \ps(c^{1/\delta}\theta, p/c) ,\quad c\geq 1.\]
So in the limit of small $p$, if $p$ is decreased by 10 dB, $\theta$ can be increased by $5\alpha$ dB to
maintain the same success probability.

\figref{fig:cell_p_contour} shows a contour plot indicating the combinations of $\theta$ and $p$ (in dB) that achieve a given target success probability $p_{\rm t}$, together with the asymptotes obtained from \eqref{limit_ps} by calculating $t$ from $t=(p_{\rm t}^{-1}-1)\sinc\delta$ and then plotting $\theta(p)=(t/p)^{1/\delta}$, which is a line in the log-log plot. Hence, keeping $p\theta^\delta$ constant results asymptotically in the
same success probability, as $p\to 0$ or $\theta\to\infty$; in contrast, in the bipolar case, keeping $p\theta^\delta$ constant results in exacty the same success probability
for all values of $p$ and $\theta$.

An important question is whether---as in the bipolar case---the variance goes to $0$ as $p\to 0$ while
keeping $\ps$ constant. The last corollary answers that question.

\begin{corollary}
Given $t=p\theta^\delta$,
\begin{equation}
\lim_{\substack{p\to 0\\\theta=(t/p)^{1/\delta}}}\var \Ps(\theta,p) =\frac{\sinc\delta}{2t+\sinc\delta} - \left(\frac{\sinc\delta}{t+\sinc\delta} \right)^2.
\end{equation}
Expressed as a function of the target success probability $p_{\rm t}$,
\begin{equation}
\lim_{\substack{p\to 0\\\theta=(t/p)^{1/\delta}}} \var \Ps(\theta,p) = \frac{p_{\rm t}}{2-p_{\rm t}}-p_{\rm t}^2.
\label{var_asymp2}
\end{equation}
\end{corollary}
\begin{IEEEproof}
The inverse of the second moment follows from Thm.~\ref{thm:cell_p} and is given by
\[ M_2^{-1}=1+2p\:\underbrace{\theta\frac{\delta}{1-\delta}\,_2F_1(1,1-\delta;2-\delta,-\theta)}_A-\:p^2\,\underbrace{\theta^2\frac{\delta}{2-\delta}\,_2F_1(2,2-\delta;3-\delta,-\theta)}_B. \]
As $\theta\to\infty$, combining \eqref{limit_ps} and \eqref{m1_p1}, $A=\theta^\delta/\sinc\delta$. For $B$,
we have\footnote{See, e.g., \url{http://dlmf.nist.gov/15.8#E2}.} $B=\Theta(\theta^{\delta})$. Hence, for some constant $c>0$,
\[ \lim_{\substack{p\to 0\\\theta=(t/p)^{1/\delta}}} M_2^{-1}=1+2t/\sinc\delta-ptc=1+2t/\sinc\delta . \]
The result follows from $\var\Ps=M_2-M_1^2$, with $M_1$ given in \eqref{limit_ps}.
\end{IEEEproof}

\begin{figure}
\centerline{\epsfig{file=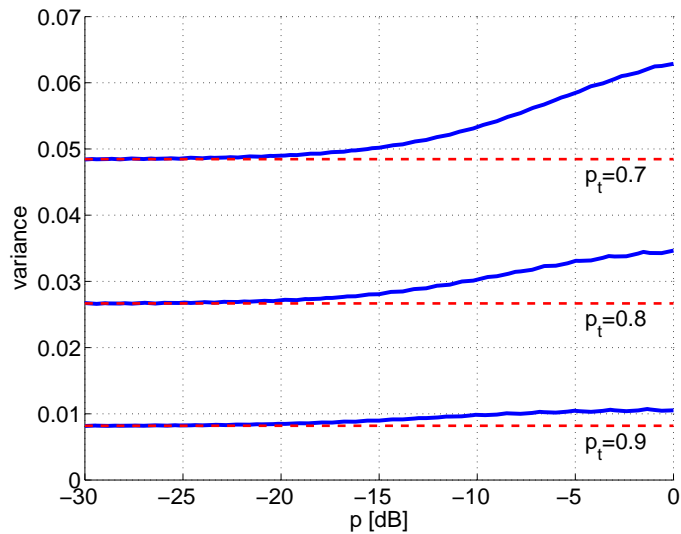,width=\figwidth}}
\caption{Variance $M_2-M_1^2$ as a function of the BS activity probability $p$ for target success probabilities
$p_{\rm t}\in\{0.7,0.8,0.9\}$ for $\alpha=4$. The dashed lines are the asymptotes from \eqref{var_asymp2}.}
\label{fig:cell_p_var}
\end{figure}

\figref{fig:cell_p_var} displays the variance as a function of $p$ for different target success probabilities. These are the variances
obtained along the corresponding contour lines in \figref{fig:cell_p_contour}. The asymptotic variance from \eqref{var_asymp2} is also shown.
It can be seen that the transmit probability has relatively little impact on the variance, especially for higher success probabilities. So, in contrast to the bipolar case, the disparity in the user experience cannot be significantly reduced by random BS activation patterns.

\section{Conclusions}
While spatial averages, such as the success probability of a transmission over
the typical link (or standard SIR distribution), are useful, they do not provide much information about the
performance of the individual links or users in a given realization of the network.
To overcome this drawback, this paper introduces the meta distribution of the SIR,
which is the distribution of the conditional SIR distribution (or success probability)
given the point process, and provides an exact expression, bounds, and an
approximation, for Poisson bipolar and cellular networks.
Hence the complete distribution of the conditional link success probability $\Ps$
in both types of Poisson  networks can be characterized.
The complete distribution of $\Ps(\theta)$ provides much more fine-grained
information that just the mean $\ps(\theta)$ that is usually consiered.

The key insight is that the moments of $\Ps$ can be calculated in closed-form. 
Hence standard and optimum moment-based bounding techniques can be
employed, which yield lower and upper bounds that are reasonably tight in some
regimes.
Moreover, an approximation by a beta distribution by matching first and second moments
turns out to be matching the exact distributions extremely accurately.

Bipolar networks with ALOHA exhibit the interesting property that the variance of $\Ps$
goes to $0$ as the transmit probability $p\to 0$ while keeping the (mean) success probability
constant. This is, however, not the case for cellular networks. If interfering base stations
are active independently with probability $p$, the variance approaches a non-zero constant
as $p\to 0$, again while keeping a constant success probability $\ps$. 
So the deployment of an ultra-dense network of small cells that are only active with small probability
(when a user requires service in their cell) does not significantly reduce the disparity of
user experiences. On the positive side,
lowering $p$ allows an increase of $\theta$ without affecting $\ps$. To be precise, decreasing
$p$ by 10 dB allows an increase of $\theta$ by $5\alpha$ dB.

From a broader perspective, the results show that it is possible in certain cases
to not only derive spatial averages, but complete {\em spatial distributions}, which
constitute rather sharp results on the network performance since they capture
the statistics of all links in a given realization of the network.
Hence it is demonstrated
that stochastic geometry allows for the calculation of (even) stronger
results than spatial averages.

\section*{Acknowledgment}
The partial support of the U.S.~National Science Foundation through grant
CCF 1216407 is gratefully acknowledged.

\appendix
\subsection{Proof of Theorem 1}
\begin{IEEEproof}
Given $\Phi$, the success probability is
\[ \Ps(\theta)=\P(h>\theta' I\mid\Phi) =\E(e^{-\theta' I}\mid \Phi),\]
where $\theta'=\theta R^\alpha$ and
\[ I=\sum_{x\in\Phi} h_x \|x\|^{-\alpha}\one(x\in\Phi_{\rm t}) .\] 
Averaging over the fading and ALOHA, it follows that
\[ \Ps(\theta)=\prod_{x\in\Phi} \frac{p}{1+\theta'\|x\|^{-\alpha}}+1-p .\]
Hence we have
\begin{align*}
  M_b&=\E\left[\prod_{x\in\Phi}\left(\frac{p}{1+\theta'\|x\|^{-\alpha}}+1-p\right)^b\right] \\
   &=\exp\left(-\lambda\int_{\R^2}\left[1-\left(\frac{p}{1+\theta'\|x\|^{-\alpha}}+1-p\right)^b\right] \dd x\right).
\end{align*}
This is the same integral as in \cite[Appendix A]{net:Haenggi13twc} and thus for $b\in\mathbb{N}$, the resulting expression
is the diversity polynomial derived there.

For general (non-integer) $b$, the proof in \cite[Appendix A]{net:Haenggi13twc} needs to be modified.
Expressing the moments as $M_b=e^{-\lambda F_b}$, we have from (29) in that paper
\[ F_b=\pi\delta\int_0^\infty \left[1-\left(1-\frac{p\theta'}{u+\theta'}\right)^b\right]u^{\delta-1} \dd u .\]

For general $b\in\mathbb{C}$, we replace the summation bound by $\infty$ since
\begin{align*}
 (1-x)^b &\equiv \sum_{k=0}^\infty \binom bk (-x)^k ,
\end{align*}
and we obtain 
\begin{align*}
  F_b&=\pi\delta\int_0^\infty \sum_{k=1}^\infty \binom bk (-1)^{k+1} (p\theta')^k \frac{u^{\delta-1}}{(u+\theta')^k}\dd u \\
   &=\pi\delta\sum_{k=1}^\infty \binom bk (-1)^{k+1} (p\theta')^k \int_0^\infty \frac{u^{\delta-1}}{(u+\theta')^k}\dd u.
 \end{align*}
 For the integral we have
 \[ \int_0^\infty \frac{u^{\delta-1}}{(u+\theta')^k}\dd u=\theta'^{\delta-k}\frac{(-1)^{k+1}\pi}{\sin(\pi\delta)}
 \frac{\Gamma(\delta)}{\Gamma(k)\Gamma(\delta-k+1)} \]
 and thus
 \begin{align*}
   F_b&=\pi\theta'^{\delta}\frac{\pi\delta}{\sin(\pi\delta)} \sum_{k=1}^\infty \binom bk p^k \frac{\Gamma(\delta)}{\Gamma(k)\Gamma(\delta-k+1)} \\ 
    &=\pi \theta^\delta R^2 \frac{\pi\delta}{\sin(\pi\delta)}  \sum_{k=1}^\infty \binom bk \binom{\delta-1}{k-1} p^k .
 \end{align*}

For the $-1$-st moment, we obtain
\[ F_{-1}=-\pi R^2\Gamma(1+\delta)\Gamma(1-\delta)\theta^\delta p (1-p)^{\delta-1},\quad p<1, \]
and thus
\begin{align*}
 M_{-1}&=\exp(C\theta^\delta p (1-p)^{\delta-1} ) \\
&= M_1^{-(1-p)^{\delta-1}},\quad p<1.
\end{align*}
\vspace*{-9.5mm}\par
\end{IEEEproof}

\bibliographystyle{IEEEtr}

\end{document}